\documentclass[twocolumn]{aastex631}

\begin{document}

\title{JWST/NIRCam Pa$\mathrm{\beta}$ Narrow-band Imaging Reveals Ordinary Dust Extinction for H$\alpha$ Emitters within the Spiderweb Protocluster at $\mathrm{z=2.16}$}

\correspondingauthor{J. M. P\'erez-Mart\'inez}
\email{jm.perez@iac.es}

\author[0000-0002-5963-6850]{Jose Manuel P\'erez-Mart\'inez}
\affiliation{Instituto de Astrof\'isica de Canarias (IAC), E-38205, La Laguna, Tenerife, Spain.}
\affiliation{Universidad de La Laguna, Dpto. Astrof\'isica, E-38206, La Laguna, Tenerife, Spain.}

\author[0000-0001-7147-3575]{Helmut Dannerbauer}
\affiliation{Instituto de Astrof\'isica de Canarias (IAC), E-38205, La Laguna, Tenerife, Spain.}
\affiliation{Universidad de La Laguna, Dpto. Astrof\'isica, E-38206, La Laguna, Tenerife, Spain.}

\author[0000-0002-0479-3699]{Yusei Koyama}
\affiliation{Department of Astronomical Science, The Graduate University for Advanced Studies, 2-21-1 Osawa, Mitaka, Tokyo 181-8588, Japan}
\affiliation{Subaru Telescope, National Astronomical Observatory of Japan, National Institutes of Natural Sciences \\
650 North A'ohoku Place, Hilo, HI 96720, USA}

\author[0000-0003-4528-5639]{Pablo G. P\'{e}rez-Gonz\'{a}lez}
\affiliation{Centro de Astrobiolog\'{\i}a (CAB), CSIC-INTA, Ctra. de Ajalvir km 4, Torrej\'{\o}n de Ardoz, E-28850, Madrid, Spain}

\author[0000-0003-4442-2750]{Rhythm Shimakawa}
\affiliation{Waseda Institute for Advanced Study (WIAS), Waseda University, 1-21-1, Nishi-Waseda, Shinjuku, Tokyo, 169-0051, Japan}
\affiliation{Center for Data Science, Waseda University, 1-6-1, Nishi-Waseda, Shinjuku, Tokyo, 169-0051, Japan}

\author[0000-0002-2993-1576]{Tadayuki Kodama}
\affiliation{Astronomical Institute, Tohoku University, 6-3, Aramaki, Aoba, Sendai, Miyagi 980-8578, Japan}

\author[0000-0001-5757-5719]{Yuheng Zhang}
\affiliation{Purple Mountain Observatory, Chinese Academy of Sciences, 10 Yuanhua Road, Nanjing, 210023, China}
\affiliation{School of Astronomy and Space Science, University of Science and Technology of China, Hefei, Anhui 230026, China}
\affiliation{Instituto de Astrof\'isica de Canarias (IAC), E-38205, La Laguna, Tenerife, Spain.}
\affiliation{Universidad de La Laguna, Dpto. Astrof\'isica, E-38206, La Laguna, Tenerife, Spain.}

\author[0000-0002-9509-2774]{Kazuki Daikuhara}
\affiliation{Astronomical Institute, Tohoku University, 6-3, Aramaki, Aoba, Sendai, Miyagi 980-8578, Japan}

\author[0000-0001-7344-3126]{Chiara D'Eugenio}
\affiliation{Instituto de Astrof\'isica de Canarias (IAC), E-38205, La Laguna, Tenerife, Spain.}
\affiliation{Universidad de La Laguna, Dpto. Astrof\'isica, E-38206, La Laguna, Tenerife, Spain.}

\author[0000-0001-7713-0434]{Abdurrahman Naufal}
\affiliation{Department of Astronomical Science, The Graduate University for Advanced Studies, 2-21-1 Osawa, Mitaka, Tokyo 181-8588, Japan}
\affiliation{National Astronomical Observatory of Japan, 2-21-1 Osawa, Mitaka, Tokyo 181-8588, Japan}




\begin{abstract}

We combine JWST/NIRCam and Subaru/MOIRCS dual Pa$\mathrm{\beta}$\,+\,H$\mathrm{\alpha}$ narrow-band imaging to trace the dust attenuation and the star-formation activities of a sample of 43 H$\mathrm{\alpha}$ emitters at the core of one of the most massive and best-studied clusters in formation at the cosmic noon: the Spiderweb protocluster at $\mathrm{z=2.16}$. We find that most H$\mathrm{\alpha}$ emitters display Pa$\mathrm{\beta}$/H$\mathrm{\alpha}$ ratios compatible with Case B recombination conditions, which translates into nebular extinction values ranging at $\mathrm{A_V\approx0-3}$ magnitudes, and dust corrected $\mathrm{Pa\beta}$ star formation rates consistent with coeval main sequence field galaxies at fixed stellar mass ($\mathrm{9.4<\log M_*/M_\odot<11.0}$) during this cosmic epoch. Furthermore, we investigate possible environmental impacts on dust extinction across the protocluster large-scale structure and find no correlation between the dustiness of its members and environmental proxies such as phase-space position, clustercentric radius, or local density. These results support the scenario for which dust production within the main galaxy population of this protocluster is driven by secular star formation activities fueled by smooth gas accretion across its large-scale structure. This downplays the role of gravitational interactions in boosting star formation and dust production within the Spiderweb protocluster, in contrast with observations in higher redshift and less evolved protocluster cores.

\end{abstract}

\keywords{Protoclusters (1297) --- Emission line galaxies(459) --- Interstellar dust extinction (837) --- Star formation (1569) --- Galaxy Evolution (594) --- Near Infrared astronomy (1093)}


\section{Introduction} \label{sec:intro}

Galaxy protoclusters are massive large-scale structures in formation residing at the nodes of the cosmic web which, upon virialization, will become the most massive and extreme density peaks at the local universe, i.e., galaxy clusters (see \citealt{Overzier16} and \citealt{Alberts22} for a review). Nevertheless, not all overdensities identified during the early universe will end up collapsing into such massive structures by $\mathrm{z=0}$, with their fate ultimately depending on their potential to accrete more matter from their surroundings and not on their cumulative mass at the time of observations (\citealt{Ata22}; \citealt{Gouin22}; \citealt{Remus23}). Thus, protoclusters are by definition extended structures, encompassing tens to hundreds of projected $\mathrm{Mpc^2}$, often composed of several local density peaks or clumps, and mainly populated by star-forming or starbursting galaxies (e.g., \citealt{Cucciati18}; \citealt{Jin21}; \citealt{Polletta21}; \citealt{Forrest23}; \citealt{Staab24}). 

Simulations predict that protoclusters host 20-50\% of the star-formation rate density of the Universe at $\mathrm{z=2-10}$ (\citealt{Chiang17}). Such vigorous star-formation can only be sustained by feeding protocluster members with cold streams of pristine gas along the filaments of the cosmic web (e.g., \citealt{Dekel06}; \citealt{Genel08}) which is crucial to explain the rapid mass build-up of the most massive galaxies at this cosmic epoch (\citealt{Daddi07}; \citealt{Bethermin14}). While part of these star-formation activities is detectable through UV-to-optical tracers (e.g., \citealt{Toshikawa16}; \citealt{Ito20}; \citealt{Newman22}, \citealt{Huang22}; \citealt{Laishram24}), a significant fraction is heavily obscured by the accompanying dust, emitting mainly in the rest-frame far infrared and submillimeter (\citealt{Smail97b}; \citealt{Greve05}). Indeed, submillimeter galaxies (also known as dusty star-forming galaxies, \citealt{Casey14}) are common tracers for high redshift overdensities (\citealt{Calvi23}). Recent studies have shown that they are the main components of protocluster cores at $\mathrm{z\gtrsim4}$ (so-called "protocores", e.g., \citealt{Oteo18}; \citealt{Long20}; \citealt{Hill22}; \citealt{Zhou24}) but they are also present across the large scale structure of more evolved and massive systems at $\mathrm{z\approx2}$ (e.g., \citealt{Smail14}; \citealt{Dannerbauer14}; \citealt{Zhang22}), with several authors proposing these objects as the progenitors of red passive ellipticals (e.g., \citealt{Lutz01}; \citealt{Swinbank06}; \citealt{Michalowski10}; \citealt{Toft14}; \citealt{Simpson14}) which dominate the cluster cores at later epochs (e.g., \citealt{Ivison13}; \citealt{Smail14}).

Nevertheless, dust attenuation hinders our ability to trace the nature of the internal physical processes such as mass growth within star-forming galaxies and their potential environmental imprints. At $\mathrm{z\lesssim2.5}$, direct measurements of dust attenuation can be achieved by comparing the observed and intrinsic flux ratios of hydrogen emission lines (i.e., hydrogen decrements), particularly using the higher transitions of the Balmer series (e.g., \citealt{Momcheva13}; \citealt{Dominguez13}; \citealt{Reddy15}; \citealt{Nelson16}; \citealt{Lorenz23}). This emission predominately originates from star-forming (H{\sc{II}}) regions in these galaxies and the expected ratios can be computed theoretically thanks to the relatively simple structure of the hydrogen atom (e.g., \citealt{Dopita03}; \citealt{Osterbrock06}). As we move closer to the cosmic noon and beyond, the strongest Balmer transitions move to the observed-frame NIR hindering their observation with ground-based facilities. The advent of the James Webb Space Telescope (JWST) allows extending the redshift range of application of this method beyond $\mathrm{z=2.5}$ (e.g., \citealt{Matharu23}; \citealt{Sandles23}). However, the star-forming activities of galaxies become more intense and so does the accompanying dust at this cosmic epoch (e.g., \citealt{PPG05}; \citealt{Madau14}; \citealt{Forster-Schreiber20}), obscuring the emission lines closer to the rest frame UV in the dustier objects. Thus, complementary dust-free star-forming tracers (e.g., Paschen series hydrogen lines in the NIR rest-frame), only accessible with JWST at $\mathrm{z>2}$, are crucial to revealing the dominant star formation mode within galaxies (i.e., starburst vs star-forming), exploring the propagation of quenching, and shedding light on the influence of early environmental effects in protoclusters.

This work will probe this technique by taking advantage of the Cycle 1 JWST/NIRCam Pa$\mathrm{\beta}$ narrow- and broad-band imaging program ID:\#1572 (PI: Helmut Dannerbauer; CoPI: Yusei Koyama) to map the star-formation activities (obscured and unobscured) and dust attenuation of H$\mathrm{\alpha}$ emitters across one of the most massive ($\mathrm{M_{500}>3.5\times10^{13}\,M_\odot}$, \citealt{DiMascolo23}) and best-studied clusters in formation at $\mathrm{z\gtrsim2}$: The Spiderweb protocluster at $\mathrm{z=2.16}$. This structure stands out above others for being the first ever reported protocluster with clear signs of a nascent ICM via Sunyaev-Zeldovich effect (\citealt{DiMascolo23}) and X-ray (\citealt{Lepore24}) implying this structure will soon become a bonafide galaxy cluster. However, it was first discovered several decades ago by \cite{Kurk00} using narrow-band imaging to identify an overdensity of Lyman-$\alpha$ emitters (LAEs) around the radio galaxy MRC1138-262 or Spiderweb galaxy (\citealt{Roettgering94}; \citealt{Pentericci97}). These initial reports were later followed up by many other spectrophotometric programs unveiling the distribution and properties of different galaxy populations including the characterization of the central radio galaxy (\citealt{Pentericci00}; \citealt{Carilli02}; \citealt{Miley06}; \citealt{Seymour07}; \citealt{Hatch08,Hatch09}; \citealt{Emonts13,Emonts16}; \citealt{DeBreuck22}; \citealt{Carilli22}; \citealt{Tozzi22b}; \citealt{Anderson22}), the study of a rich population of emission line galaxies including H$\alpha$ emitters (HAEs) which form the backbone of the protocluster population (\citealt{Kurk04b,Kurk04a}; \citealt{Kuiper11}; \citealt{Hatch11}; \citealt{Koyama13}; \citealt{Shimakawa14, Shimakawa15, Shimakawa18b}; \citealt{PerezMartinez23}; \citealt{Naufal23}; \citealt{Daikuhara24}), the location of several AGN candidates throughout X-ray emission (\citealt{Pentericci02}; \citealt{Croft05}; \citealt{Tozzi22a}; \citealt{Shimakawa24}; \citealt{Lepore24}), the observation of an emerging red sequence within the cluster core (\citealt{Kurk04a}; \citealt{Kodama07}; \citealt{Zirm08}; \citealt{Doherty10}; \citealt{Tanaka10, Tanaka13}), the discovery of a network of FIR Herschel sources (\citealt{Rigby14}), starbursty submillimeter galaxies (SMGs, \citealt{Dannerbauer14}; \citealt{Zeballos18}) and gas-rich CO emitters (\citealt{Dannerbauer17}; \citealt{Emonts18}; \citealt{Tadaki19}; \citealt{ChenZ24}) including  the mapping of the protocluster large scale structure in CO(1-0) with the Australian Telescope Compact Array (ATCA, \citealt{Jin21}). This wealth of multiwavelength data makes the Spiderweb protocluster a legacy field for studying galaxy evolution in overdense environments at $\mathrm{z>2}$.

This manuscript will focus on characterizing the attenuation of HAEs in the Spiderweb field and searching for potential correlations between the dusty nature of our sources and environmental proxies within this large-scale structure. We refer the reader to the companion paper (Shimakawa et al. 2024c, hereafter S24) for a full description of the selection of Pa$\beta$ emitters (PBEs). This work is organized in the following way: $\text{Sect.}$\,\ref{S:OBs} describes our new JWST/NIRCam observations as well as the ancillary narrow-band datasets targeting H$\mathrm{\alpha}$ used for our analyses. $\text{Sect.}$\,\ref{S:Methods} describe the methods used to extract and correct our sample's final emission line fluxes and outline the environmental proxies that will be used later in $\text{Sect.}$\,\ref{S:Results} to present and discuss our results.  Finally, $\text{Sect.}$\,\ref{S:Conclusions} summarizes the major conclusions of this manuscript. Throughout this work, we assume a \citet{Chabrier03} initial mass function (IMF), and adopt a flat cosmology following the WMAP nine-year data (\citealt{Hinshaw13}) with $\Omega_{\Lambda}$=0.714, $\Omega_{m}$=0.286, and $\mathrm{H_{0}}$=69.3 km\,s$^{-1}$Mpc$^{-1}$. All magnitudes quoted in this paper are in the AB system (\citealt{Oke83}).

\section{Datasets and sample selection}
\label{S:OBs}

\subsection{JWST/NIRCam imaging}
\label{SS:NIRCam} 
This study makes use of NIR imaging data in the Spiderweb protocluster from the JWST/NIRCam Cycle 1 program ID \#1572, taken in filters F115W, F182M, F410M, and F405N. Details about the observations and data reduction can be found in the companion paper S24. This work will primarily use the JWST/NIRCam F405N narrow-band filter which captures the $\mathrm{Pa\beta}$ emission line at $\mathrm{z\approx2.16}$, and the F410M medium band filter to subtract the continuum around $\mathrm{Pa\beta}$ within our sources (see Sect.\,\ref{SS:EL}). The integration time for these two filters amounts to 1.05 and 0.36 hours respectively. Given the proximity of their central wavelength, the empirical PSF in both filters is very similar yielding $\mathrm{FWHM\approx0.14\arcsec}$ and a pixel scale of $\mathrm{0.03\arcsec}$. The 5$\sigma$ average limiting magnitude is 23.3 mag for F405N and 24.6 mag for F410M which allows tracing star formation in galaxies down to $\mathrm{SFR(Pa\beta)\approx4\,M_\odot/yr}$ assuming the \cite{Kennicutt98} calibration adjusted for a \cite{Chabrier03} IMF, and a dustless ratio of $\mathrm{H\alpha/Pa\beta=17.6}$ (\citealt{Osterbrock06}).

\subsection{Sample selection}
\label{SS:SS}

\begin{figure}
 \centering
      \includegraphics[width=\linewidth]{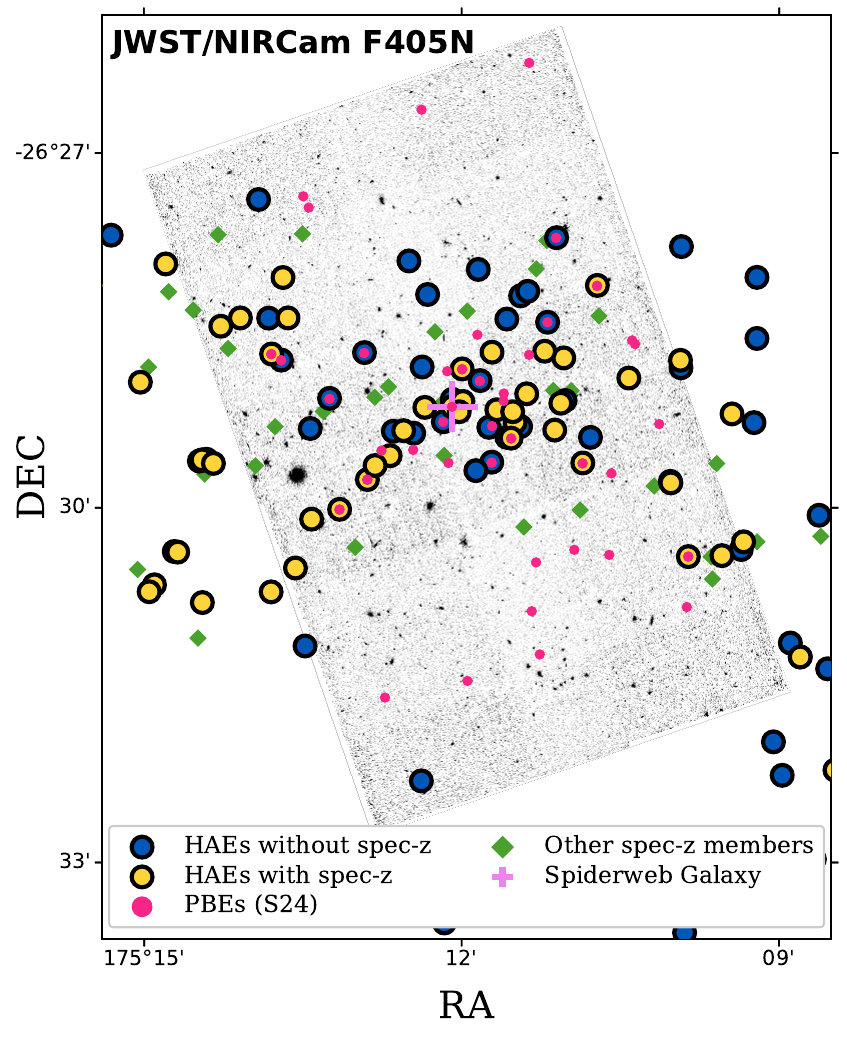}\par 
      \caption{Source distribution at $\mathrm{2.12<z<2.21}$ across the Spiderweb protocluster over the JWST/NIRCam F405N image. Magenta circles display the 42 color-selected narrow-band PBEs (S24). Big yellow and blue circles respectively depict the known HAEs with and without spectroscopic redshift (\citealt{Kurk04b,Kurk04a}, \citealt{Koyama13}; \citealt{PerezMartinez23}; \citealt{Shimakawa24}). Green diamonds show additional spectroscopic samples in this field (\citealt{Pentericci00}; \citealt{Croft05}; \citealt{Kuiper11}; \citealt{Emonts18}; \citealt{Tadaki19}; \citealt{Jin21}). The violet cross marks the position of the Spiderweb radio galaxy (\citealt{Roettgering94}). 
      }
      \label{F:Map}
      \end{figure}

More than twenty years ago, \cite{Kurk04b,Kurk04a} carried out the first attempts to identify HAEs within the Spiderweb protocluster field and spectroscopically confirm their membership. Later on, the MAHALO-Subaru project (\citealt{Kodama13}) expanded these seminal works by carrying out multiple narrow-band campaigns with Subaru/MOIRCS (\citealt{Ichikawa06}; \citealt{Suzuki08}) aimed at detecting H$\alpha$ emissions at $\mathrm{z\approx2.16}$ using the NB2071 filter in combination with $\mathrm{K_s}$ broad-band imaging. As a consequence, more than 100 HAEs have been unveiled through this technique (\citealt{Koyama13}; \citealt{Shimakawa18b}), with $\mathrm{>60}$ of them being spectroscopically confirmed protocluster members in subsequent follow-up studies (\citealt{Shimakawa15}; \citealt{PerezMartinez23}). This work will revisit these datasets to remeasure the H$\mathrm{\alpha}$ flux of the 58 known HAEs within the new JWST/NIRCam surveyed area. These HAEs will be the primary target of this work and their measured positions will act as a proxy to measuring their Pa$\mathrm{\beta}$ flux. Fig.\,\ref{F:Map} displays the spatial distribution of HAEs with and without spectroscopic redshift over the JWST/NIRCam image. Other populations of interest such as the PBEs (see S24), ancillary spectroscopic members, and the position of the Spiderweb galaxy are also shown in this figure for reference.
%
The need for measuring the H$\mathrm{\alpha}$ and Pa$\mathrm{\beta}$ emission line simultaneously restricts our field to an area equivalent to 5.6\,$\mathrm{Mpc^2}$ excluding only the upper right corner of our 6.4\,$\mathrm{Mpc^2}$ JWST/NIRCam field (Fig.\,\ref{F:Map}) which is not covered by the ancillary MAHALO-Subaru 12.8\,$\mathrm{Mpc^2}$ fields or spectroscopic follow-ups. However, this corner was covered by FIR and submillimeter programs in this field (e.g., \citealt{Rigby14}; \citealt{Dannerbauer14}; \citealt{Zeballos18}). Based on earlier works, we locate 58 HAEs within this field split into two subsets: 33 HAEs have spectroscopic redshift and thus they are confirmed protocluster members, while 25 of them lack this information (yellow and blue circles in Fig.\,\ref{F:Map} respectively). These two subsets represent our primary sample and will be quoted in this way both in the subsequent figures and the text. 

The transmission curves of the SUBARU/MOIRCS NB2071 and JWST/NIRCam F405N filters do not perfectly match for the simultaneous detection of the H$\mathrm{\alpha}$ and Pa$\mathrm{\beta}$ emission lines and thus, we need to remove spectroscopic sources for which any of these two emission lines lie near the edges of their respective bandwidths. As a consequence, we exclude two spectroscopic sources that lie at $<10\%$ of the maximum throughput of either the NB2071 or the F405N filters. Out of the remaining 56 HAEs, 17 sources are classified as PBEs in S24. In short, these sources fulfill that $\mathrm{EW(Pa\beta)_{rest}>20\,\AA}$ and display F410M-F405N detection at $\mathrm{>3\sigma}$. We refer the reader to S24 where extensive details about the selection criteria and PBE definition can be found. Nevertheless, we expect that most HAEs have detectable levels of Pa$\beta$ emission (see Sect.\,\ref{SS:SS}) given the spectroscopically confirmed star-forming nature of these sources in previous publications (\citealt{Shimakawa15}; \citealt{PerezMartinez23}) and the unobscured star formation rate limit of our observations, $\mathrm{SFR(Pa\beta)\approx4\,M_\odot/yr}$, assuming the \cite{Kennicutt98} $\mathrm{H\alpha}$ star-forming law and a dust-free H$\mathrm{\alpha}$/Pa$\mathrm{\beta}$ flux ratio of 17.6 (\citealt{Osterbrock06}). This threshold in star formation approximately matches the narrow-band $\mathrm{H\alpha}$ limits ($\mathrm{SFR(H\alpha)\approx5\,M_\odot/yr}$) reported in \cite{Shimakawa18b}, and it is representative of main sequence galaxies down to $\mathrm{M_*\approx10^9\,M_\odot}$ at $\mathrm{z=2.16}$ (\citealt{Speagle14}).

\section{Methods}
\label{S:Methods}

\subsection{Image homogenization}
\label{SS:ImageMatch}

The main goal of this work is to measure the attenuation of narrow-band selected protocluster HAEs using the H$\mathrm{\alpha}$/Pa$\mathrm{\beta}$ flux ratio as a proxy. To achieve this, we must ensure that both the H$\mathrm{\alpha}$ and Pa$\mathrm{\beta}$ emission line fluxes are measured consistently considering the different instrumental and observational conditions. This requires the pixel and PSF homogenization of the NIRCam and MOIRCS imaging. The empirical PSF size of JWST/NIRCam observations varies from $\mathrm{0.04\arcsec}$ in the F115W filter to $\mathrm{0.137\arcsec}$ in the F410M band with a constant pixel scale of $\mathrm{0.03\arcsec/pix}$. These values are considerably smaller than the PSF ($\mathrm{\sim0.7\arcsec}$) and pixel size ($\mathrm{0.117\arcsec/pix}$) achieved by the Subaru/MOIRCS ground-based observations (\citealt{Shimakawa18b}). We choose to degrade the NIRCam image quality to match the values of the MOIRCS program while avoiding flux losses through the following steps: First, we run {\sc{SExtractor}} (\citealt{Bertin96}) to measure single band photometry and structural parameters of all sources within the field similarly to S24. We feed these measurements to the software {\sc{PSFEx}} (\citealt{Bertin11}) which selects stars with sufficient S/N within the field and creates an average PSF for each NIRCam band. We repeat this process over the Subaru/MOIRCS NB2071 imaging and use a self-written {\sc{Python}} script including the {\sc{Astropy}} and {\sc{Photutils}} libraries (\citealt{Astropy13, Astropy18}; \citealt{Photutils23}) to create a matching kernel between the ground- and space-based images. Finally, we apply this kernel to smooth the NIRCam images achieving approximately the same PSF size as Subaru/MOIRCS, and tailor them to the same dimensions and world coordinate reference frame. At this point, we run {\sc{SExtractor}} again using the Subaru/MOIRCS NB2071 image as a reference for source detection and perform dual-band photometry using flexible (i.e., KRON) apertures over the JWST/NIRCam F405N and F410M bands and the Subaru/MOIRCS NB2071 and Ks bands. Contamination from nearby detected objects is subsequently identified and subtracted during this process. 

\begin{figure*}
\centering
\includegraphics[width=0.5\linewidth]{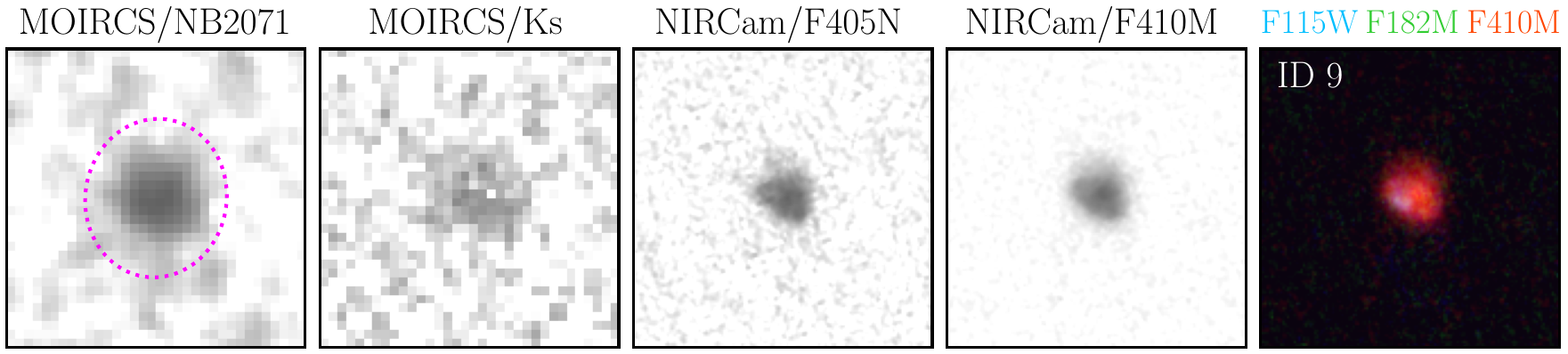}~
\includegraphics[width=0.5\linewidth]{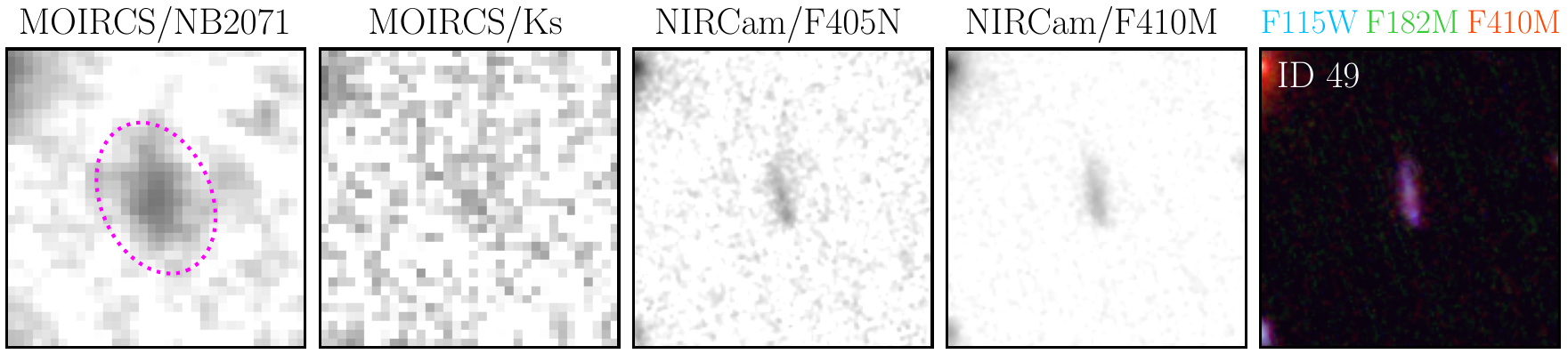}\par
\includegraphics[width=0.5\linewidth]{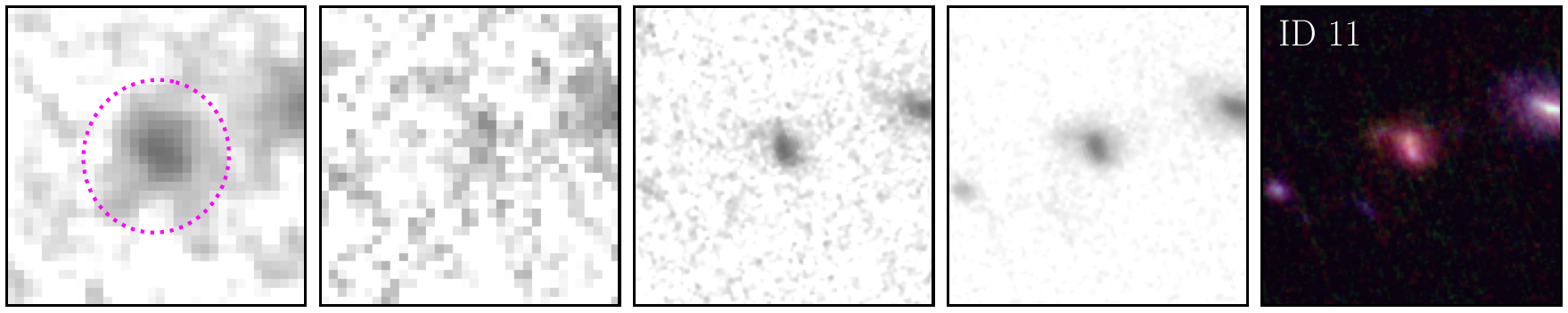}~
\includegraphics[width=0.5\linewidth]{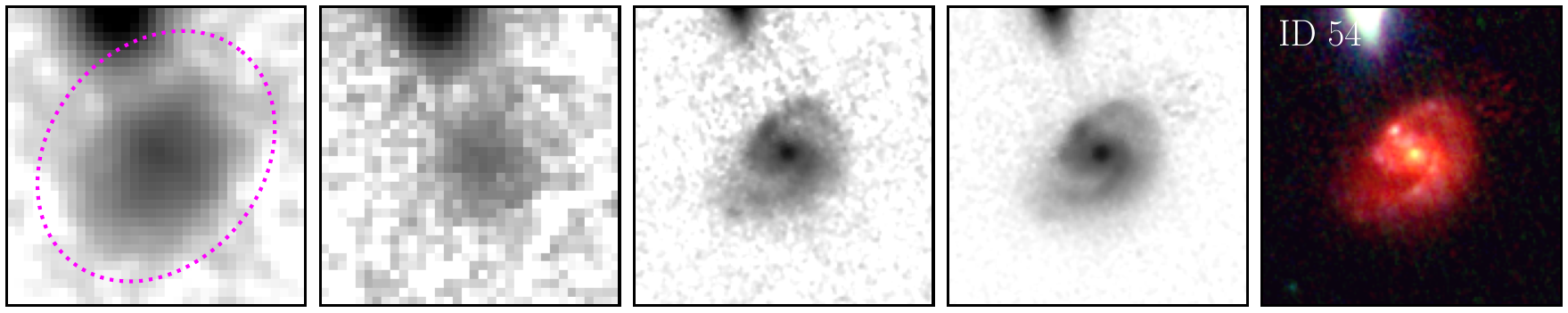}\par
\includegraphics[width=0.5\linewidth]{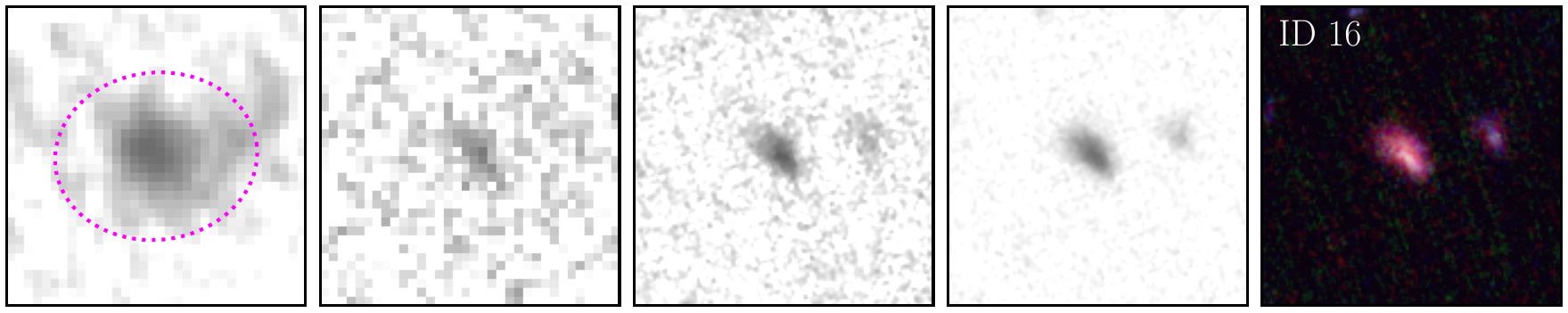}~
\includegraphics[width=0.5\linewidth]{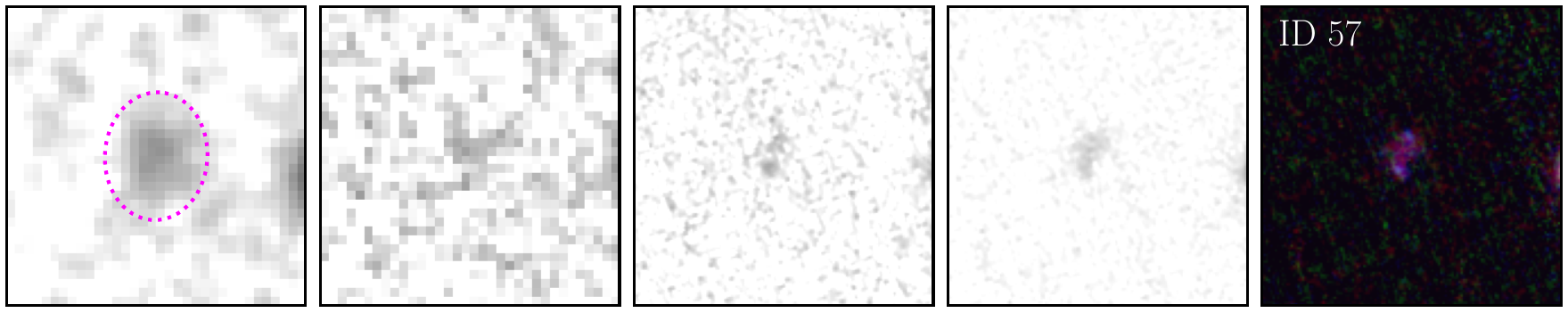}\par
\includegraphics[width=0.5\linewidth]{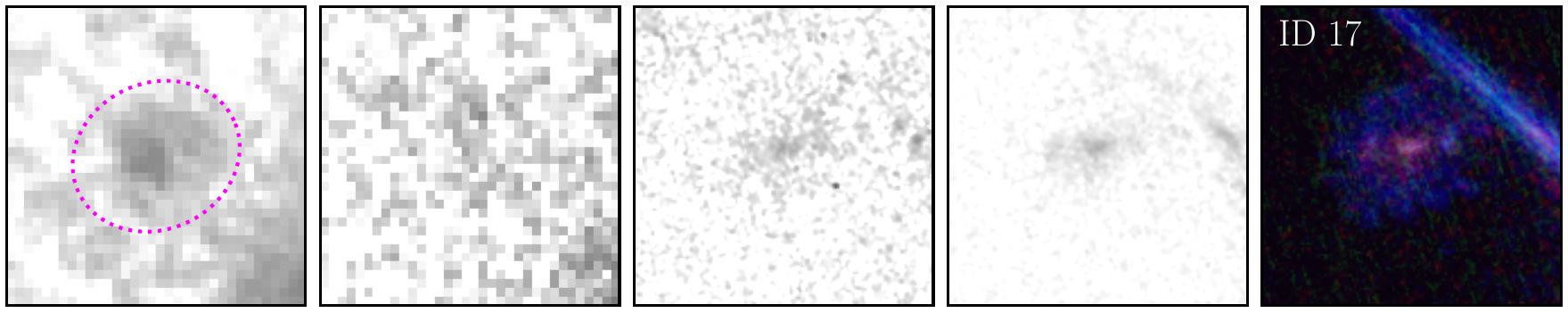}~
\includegraphics[width=0.5\linewidth]{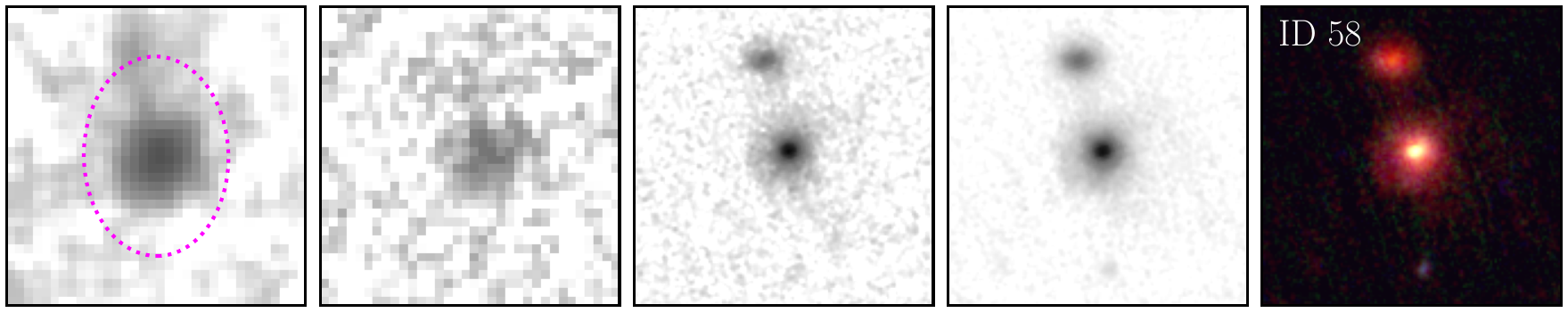}\par
\includegraphics[width=0.5\linewidth]{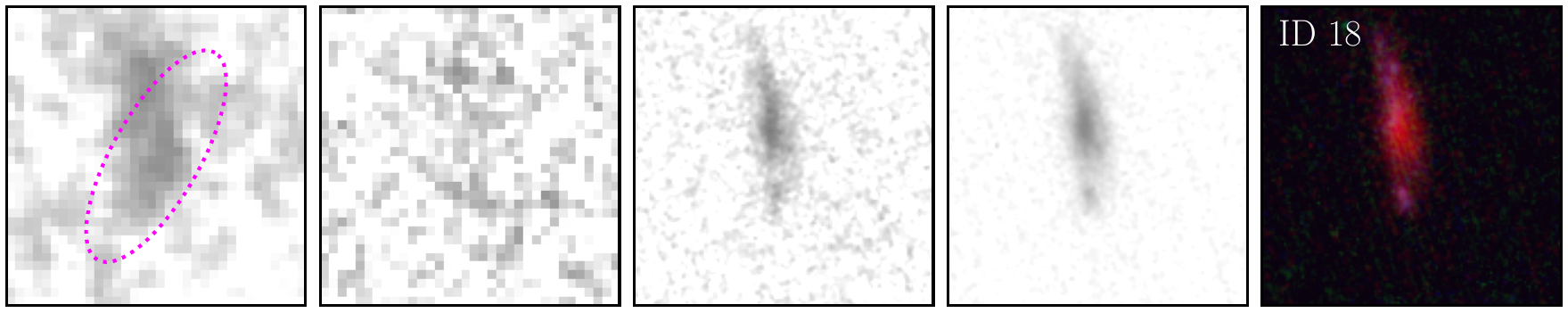}~
\includegraphics[width=0.5\linewidth]{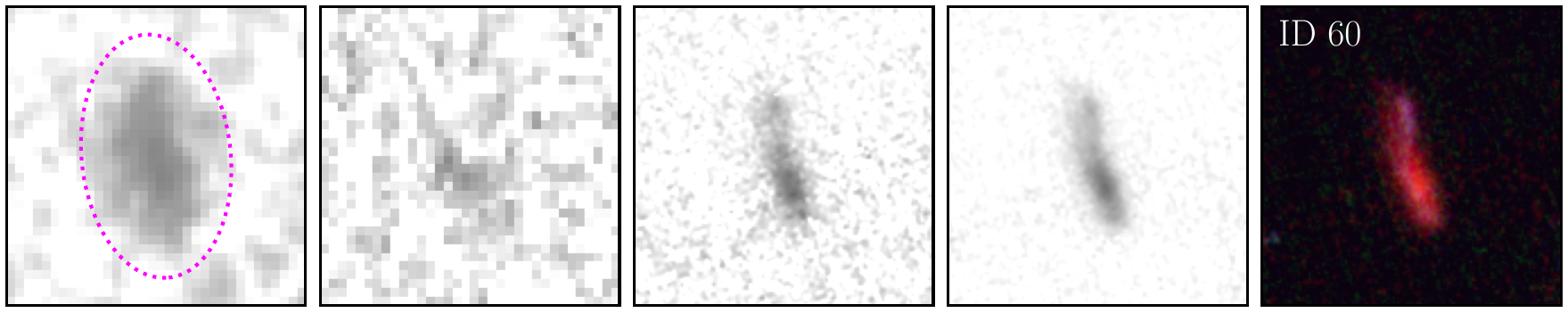}\par
\includegraphics[width=0.5\linewidth]{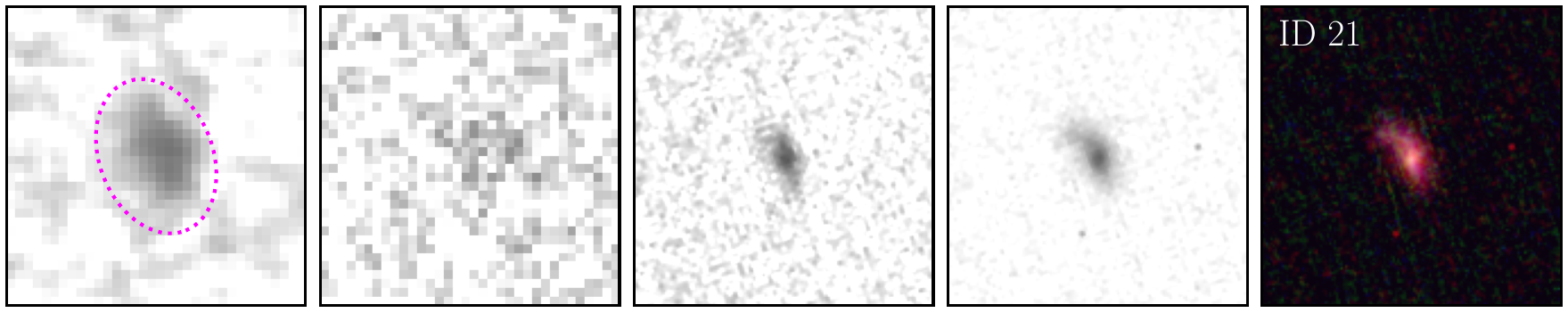}~
\includegraphics[width=0.5\linewidth]{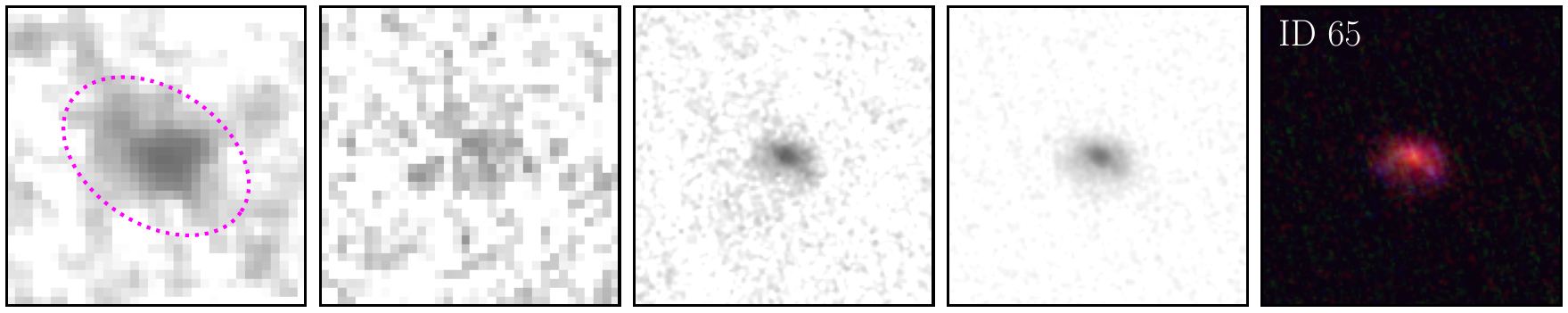}\par
\includegraphics[width=0.5\linewidth]{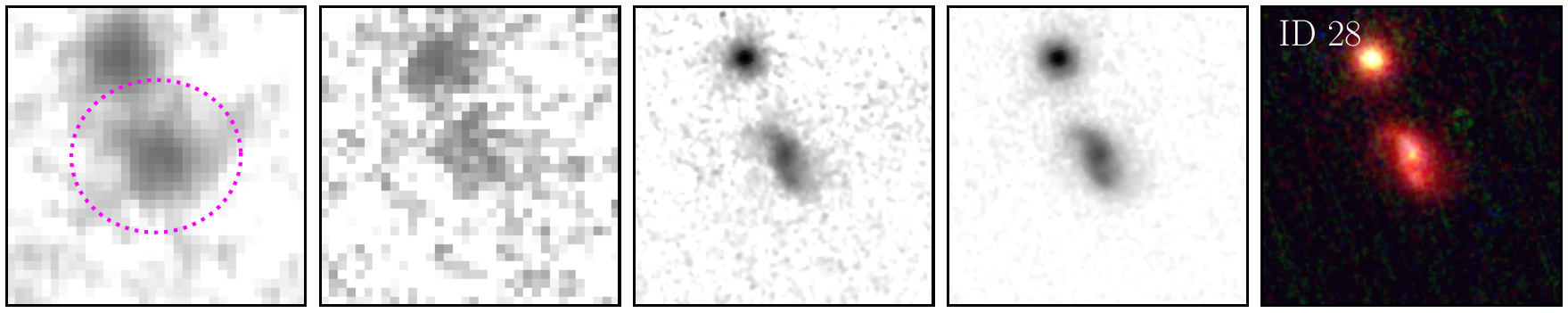}~
\includegraphics[width=0.5\linewidth]{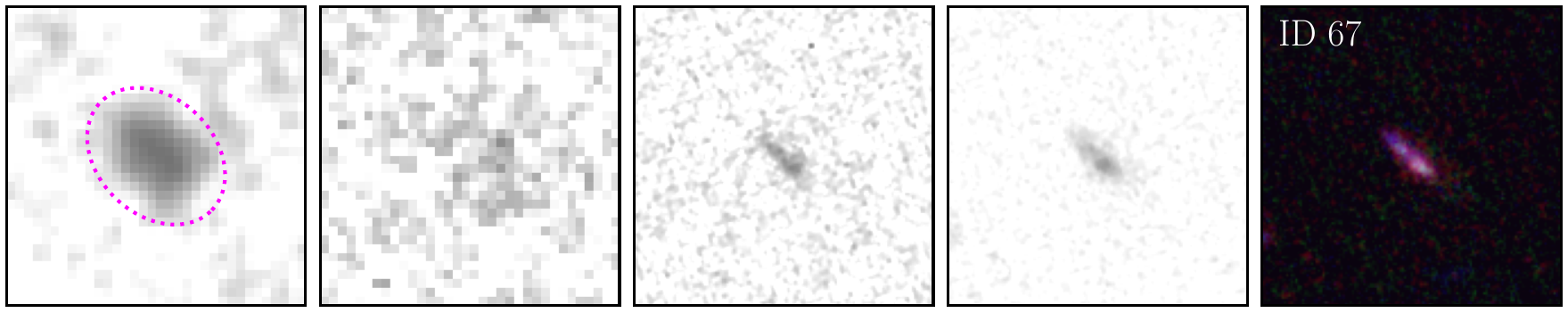}\par
\includegraphics[width=0.5\linewidth]{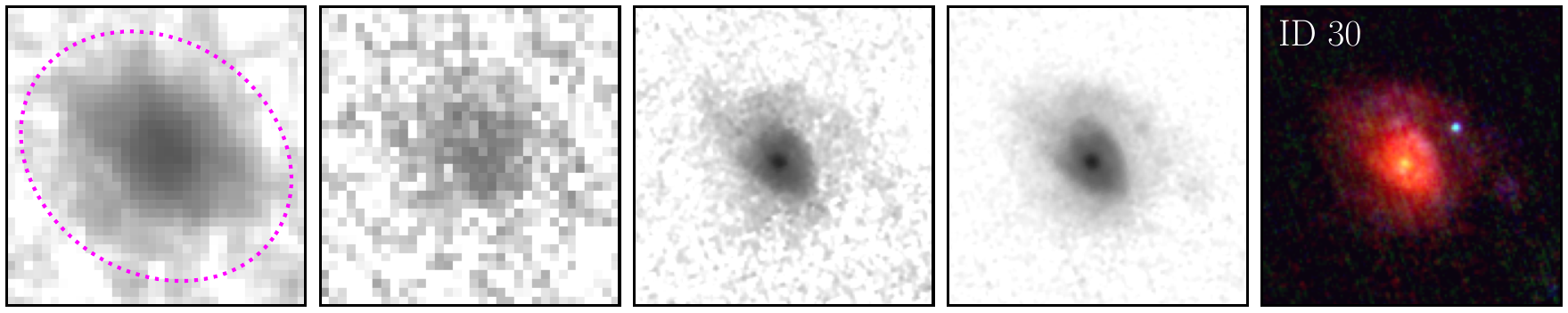}~
\includegraphics[width=0.5\linewidth]{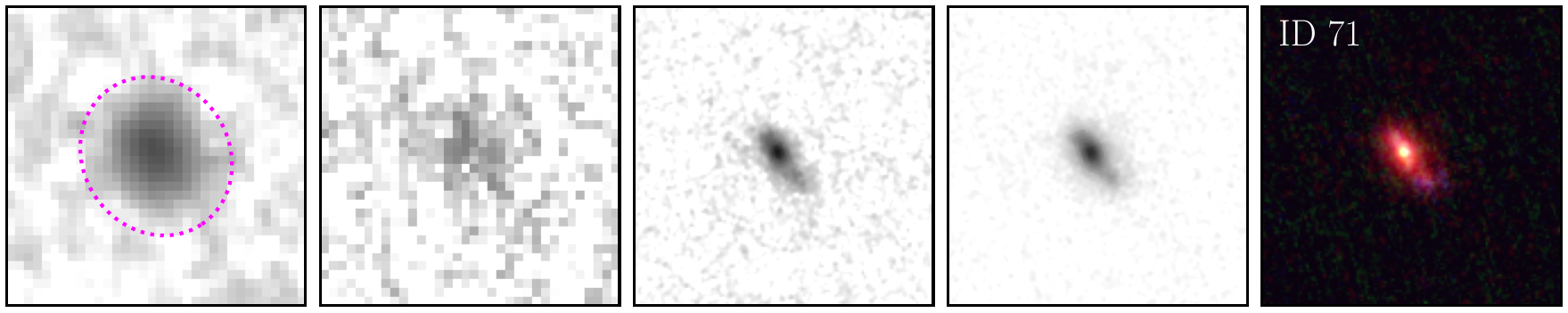}\par
\includegraphics[width=0.5\linewidth]{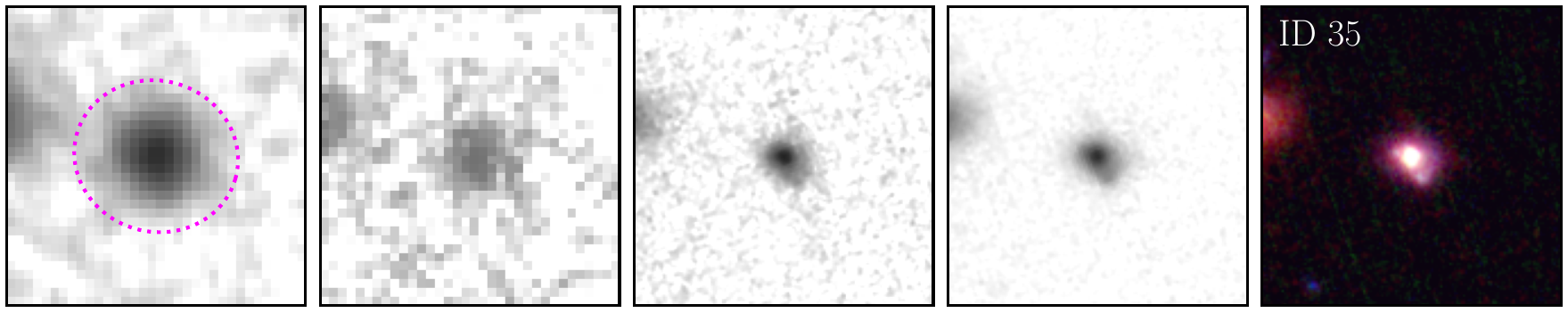}~
\includegraphics[width=0.5\linewidth]{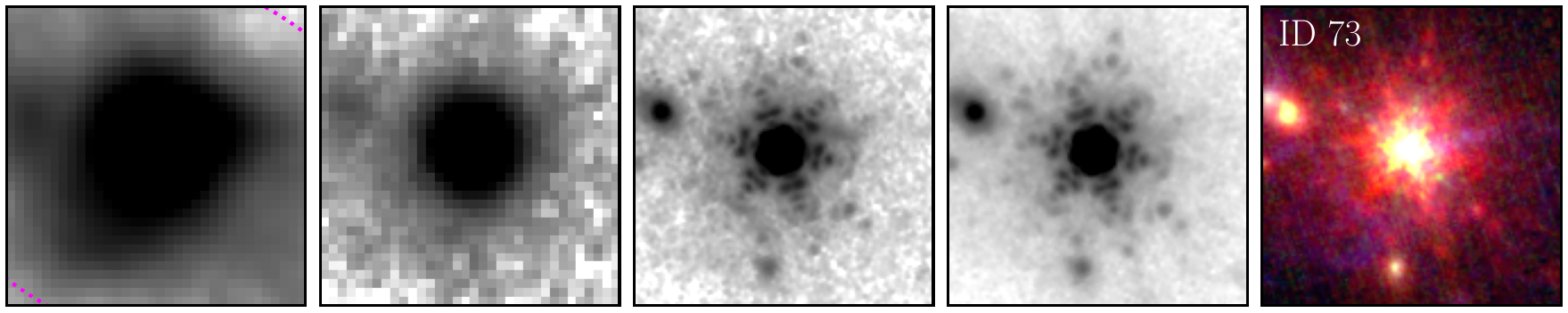}\par
\includegraphics[width=0.5\linewidth]{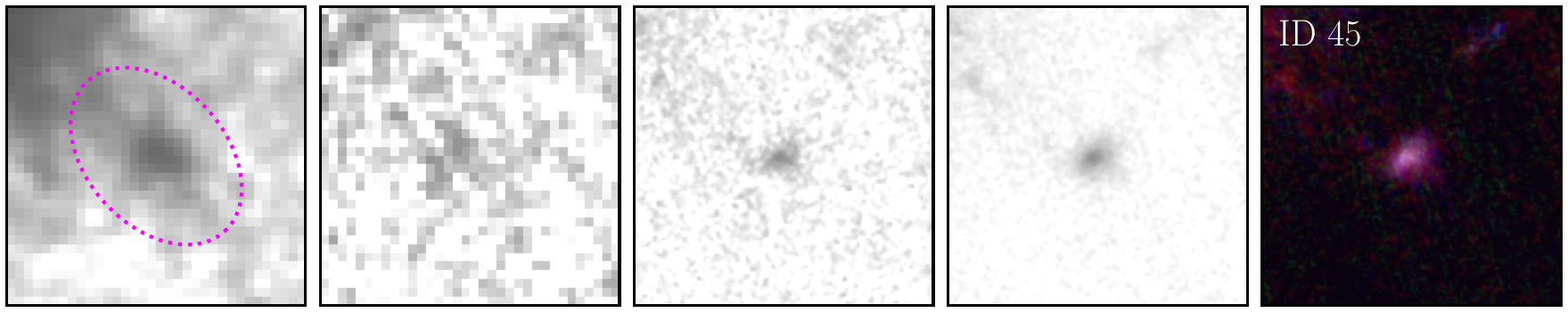}~
\includegraphics[width=0.5\linewidth]{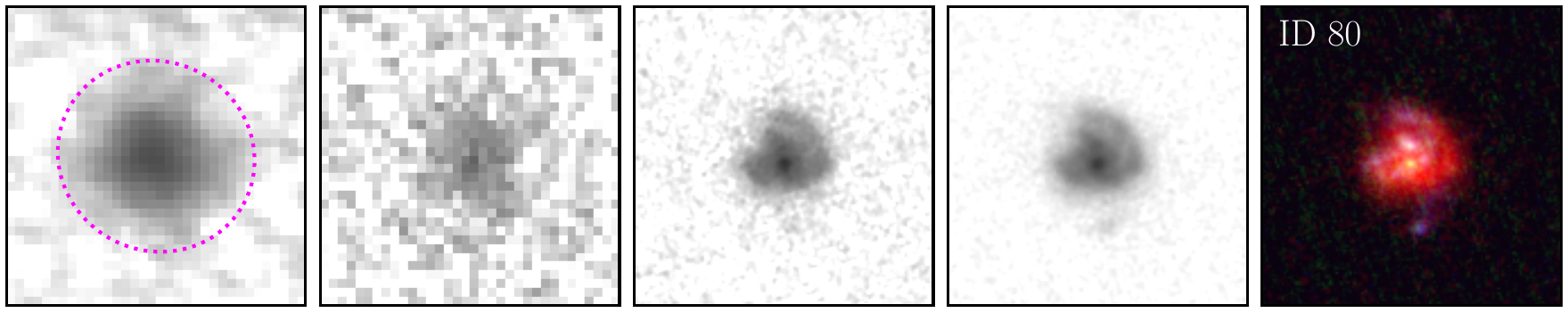}\par
\includegraphics[width=0.5\linewidth]{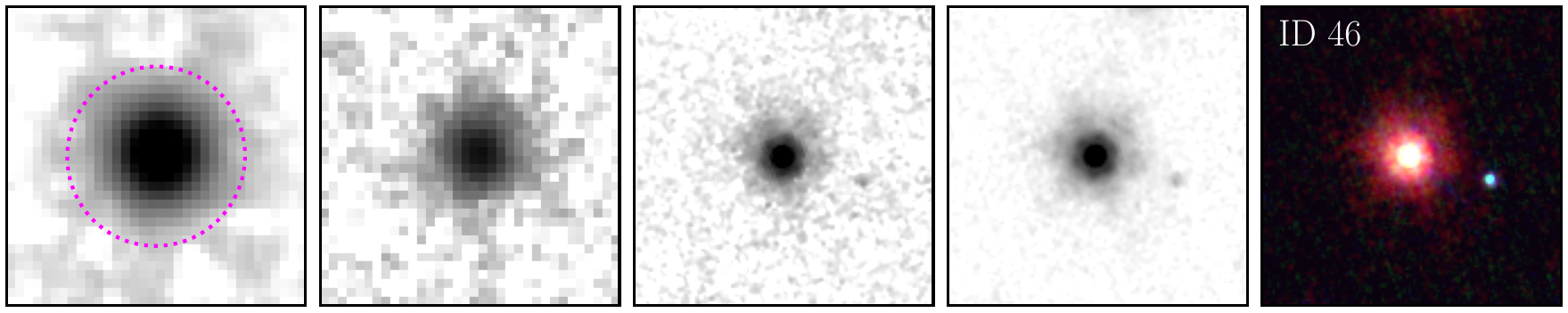}~
\includegraphics[width=0.5\linewidth]{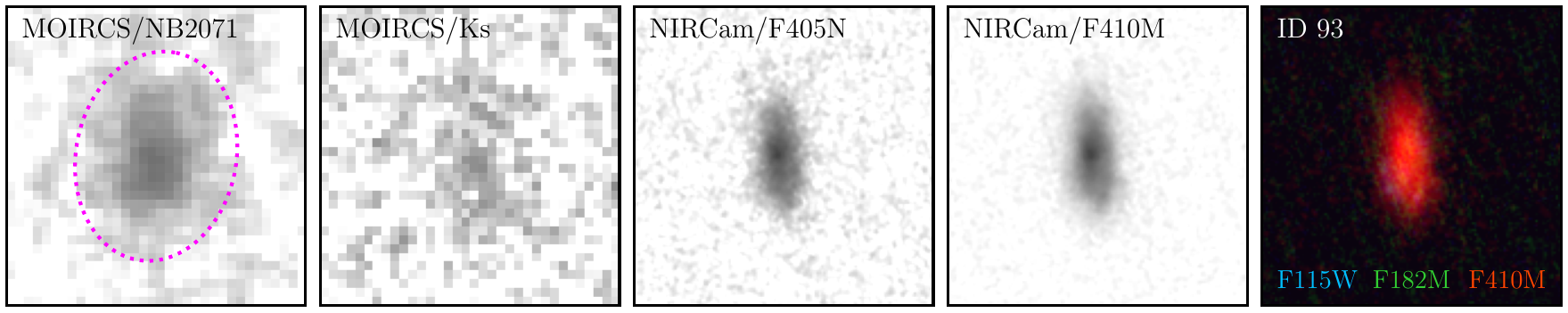}\par
\caption{Gallery depicting the images of the HAEs with spectroscopic redshift quoted in Table\,\ref{T:appendix}. Sources are ordered by the ID given in \cite{Shimakawa24} and organized in two blocks made of 5 columns each displaying squared cutouts ($\mathrm{4\arcsec\times4\arcsec}$) for each HAE.  The cutouts respectively correspond to the SUBARU/MOIRCS NB2071 and Ks filters, the JWST/NIRCam F405N and F410M filters, and a stacked RGB color that combines the JWST/NIRCam F115W, F182M, and F410M broad and medium band filters, roughly matching the rest-frame U, V and J bands. The first cutout column of each block includes a magenta dotted line displaying the size of the aperture used to extract the flux in the four presented filters (see Sect.\,\ref{SS:EL}). ID 73 corresponds with the Spiderweb galaxy (\citealt{Miley06}).}
\label{F:Gallery1}
\end{figure*}

This approach ensures that we are measuring the flux of each source in a consistent way between instruments and filters, which is crucial to obtain precise H$\mathrm{\alpha}$/Pa$\mathrm{\beta}$ emission line flux ratios. Fig.\,\ref{F:Gallery1} display $\mathrm{4\arcsec\times4\arcsec}$ cutouts for the two MOIRCS and two NIRCam original images centered over our final sample of spectroscopic HAEs and following the galaxy IDs quoted in \cite{Shimakawa24}. For reference, we display the KRON aperture (dotted magenta line) used to measure its flux in all passbands. Finally, we display a stacked RGB color image that combines the JWST/NIRCam F115W, F182M, and F410M broad and medium band filters of this program, roughly matching the rest-frame U, V, and J bands. A gallery for the remaining objects included of our final sample can be found in Fig.\,\ref{F:Gallery2}. Finally, we note that this work focuses only on obtaining Pa$\mathrm{\beta}$ fluxes for the sample of known HAEs (\citealt{Shimakawa24}) overlapping with the JWST footprint presented in Sect.\,\ref{S:OBs} and thus, we refer to S24 for an in-depth discussion on the detection and reliability of PBEs lacking HAEs counterparts (see Fig.\,\ref{F:Map}).

\subsection{H$\mathrm{\alpha}$ and Pa$\mathrm{\beta}$ emission line flux}
\label{SS:EL}

The narrow-band technique has been widely used in the past to select emission-line galaxies in relatively narrow redshift slices within a given field (e.g., \citealt{Bunker95}; \citealt{Malkan96}; \citealt{PPG03}). Subsequent spectroscopic follow-ups have demonstrated the high success rate ($\mathrm{\gtrsim90\%}$, e.g., \citealt{Sobral13}; \citealt{Shimakawa18b}; \citealt{PerezMartinez23}) of this technique when strong emission lines such as $\mathrm{H\alpha}$ are targeted in combination with color-color selection methods. This allows the identification and removal of fore-/background contaminants both in the field as well as in (proto-)clusters. Furthermore, combining both broad-band (BB) and narrow-band (NB) imaging allows the isolation of the emission line flux contribution ($\mathrm{F_{line}}$) from the underlying continuum. The exact expressions to compute the emission line flux have been described in previous works (e.g., \citealt{Koyama13}), which we reproduce here for the reader's convenience:
\begin{equation}
\label{EQ:NB}
F_{\mathrm{line}} = \frac{f_{\mathrm{NB}}-f_{\mathrm{BB}}}{1-\Delta_{\mathrm{NB}}/\Delta_{\mathrm{BB}}}\Delta_{\mathrm{NB}}
\end{equation}
with $f_{\mathrm{NB}}$ and $f_{\mathrm{BB}}$ being the narrow- and broad-band flux densities respectively, and $\Delta_{\mathrm{NB}}$ and $\Delta_{\mathrm{BB}}$ being the bandwidth of the NB and BB filters. This equation is applied to the pair of JWST/NIRCam F405N and F410M imaging to obtain the Pa$\mathrm{\beta}$ emission line flux, and to the Subaru/MOIRCS NB2071 and $\mathrm{K_{s}}$ imaging to obtain the H$\mathrm{\alpha}$ emission line flux. In the latter case, however, the recovered H$\mathrm{\alpha}$ flux is contaminated by the neighboring [N{\sc{ii}}]$\lambda\lambda$6548,6584 doublet. We estimate the amount of [N{\sc{ii}}] contamination using the [N{\sc{ii}}]/H$\mathrm{\alpha}$-based average mass-metallicity relation of HAEs in the Spiderweb protocluster computed by \cite{PerezMartinez23}, the fixed [N{\sc{ii}}] doublet flux ratio $F_{\lambda6584}/F_{\lambda6548}=3$ (\citealt{Storey00}), and the stellar mass values derived in \cite{Shimakawa24} for these sources. In that work, the stellar mass values were computed through SED modeling with X-Cigale (\citealt{Boquien19}; \citealt{Yang20,Yang22}) using multiwavelength coverage from the X-rays to $\mathrm{850\mu m}$ and including the module dedicated to the treatment of AGN activity. We refer to Sect. 3.2 in \cite{Shimakawa24} for further details about this procedure. 

The number of PBEs, as defined by S24, is relatively small within the HAEs sample, with only 17 sources simultaneously identified as HAEs and PBEs. This situation arises despite matching our JWST/NIRCam F405N program to the H$\mathrm{\alpha}$ flux limit achieved by the MAHALO-Subaru campaigns (see Sect.\,\ref{SS:SS}). This apparent contradiction is caused by the use of small apertures (0.3\arcsec) for the PBEs selection. This approach maximizes the sensitivity of the selection, thus capturing the faintest sources with relatively high S/N. However, it also misses the Pa$\mathrm{\beta}$ flux originating in the outskirts of the most massive ($\mathrm{M_*\gtrsim10^{10}\,M_\odot}$) and spatially extended HAEs (see S24 for further details). In this work, however, we use flexible (KRON) apertures based on the source detection performed over the Subaru/MOIRCS NB2071 filter, which recovers $\mathrm{Pa\beta}$ fluxes for most sources. After careful analysis, we identified 13 HAEs that show no signs of Pa$\mathrm{\beta}$ emission after applying Eq.\,\ref{EQ:NB}. Six of them are spectroscopically confirmed cluster members while the remaining seven are only detected throughout Subaru/MOIRCS narrow-band H$\mathrm{\alpha}$ imaging. 

Of these objects, two HAEs with spectroscopic redshift are classified as X-ray emitters in \cite{Tozzi22a}, and thus, they are likely AGNs. One of them was originally identified as an X-ray emitter (ID X5) with spectroscopically confirmed Ly$\mathrm{\alpha}$ emission by \cite{Croft05}, and also classified as an extremely red object (ERO) in \cite{Kurk04a}. It is a possibility that the SUBARU/MOIRCS narrow-band emission being dominated by {\sc{[nii]}} and not H$\mathrm{\alpha}$, which would explain the lack of measurable $\mathrm{Pa\beta}$ emission albeit we currently lack spectroscopic information in the required wavelength range to confirm this scenario. The other source was spectroscopically confirmed by \cite{PerezMartinez23} using VLT/KMOS (ID 880) and displays a very compact but faint emission at the wavelength of H$\mathrm{\alpha}$+{\sc{[nii]}}. Furthermore, this source lies $\mathrm{>1}$ dex below the main sequence in \citealt{Shimakawa24} as measured by multi-wavelength SED fitting, and thus can be classified as a near quiescent object. This would explain why we do not detect significant Pa$\mathrm{\beta}$ emission. In addition, as a consequence of choosing the SUBARU/MOIRCS NB2071 filter ($\mathrm{FWHM\approx0.7\arcsec}$) as the reference image for SExtractor dual-imaging mode (see Sect.\,\ref{SS:ImageMatch}), the typical KRON aperture diameter within our sample varies between $\mathrm{1\arcsec}$ and $\mathrm{2\arcsec}$. This range is significantly larger than the apparent JWST/NIRCam size of most of the remaining 11 discarded sources, which display very compact emission in F405N and F410M ($\mathrm{\lesssim0.5\arcsec}$). Thus, by degrading the JWST/NIRCam image quality to that of SUBARU/MOIRCS, the compact light profiles of these objects would have been spread over a larger area, decreasing their S/N and potentially washing out their $\mathrm{Pa\beta}$ emission. For comparison purposes, we display the imaging cutouts of these objects in Fig.\,\ref{F:Gallery3}. 

Thus, the number of galaxies with measurable $\mathrm{H\alpha+Pa\beta}$ emission is reduced to 25 HAEs with spectroscopic redshift and 18 HAEs solely identified through narrow-band imaging. For the spectroscopic sample, we compute the exact wavelength of the H$\mathrm{\alpha}$ and Pa$\mathrm{\beta}$ emission lines and correct their narrow-band fluxes from potential transmission losses caused by the differences between the two narrow-band filters' throughput curves. Given the redshift distribution of our sample, this correction is kept within 5\% for the vast majority of our objects. Finally, we abstain from attempting further corrections to the sample lacking spectroscopic redshift information.

\subsection{Dust extinction from hydrogen decrements}
\label{SS:DustExt}

Dust extinction can be computed from a diverse set of observables at different wavelength ranges. In this work, we use the so-called hydrogen decrements, i.e., ratios between hydrogen emission lines that originate within the same H{\sc{ii}} region. The photoionization of hydrogen atoms and their posterior recombination into emission lines is caused by the radiation field of hot young stars within star-forming regions, with collisional excitation being a subdominant process. The expected values of these decrements can be constrained thanks to the simple atomic structure of the hydrogen atom and it only depends on a few ISM parameters such as the electron density ($\mathrm{n_e}$) and the gas temperature ($\mathrm{T_e}$). Assuming the case B recombination, i.e., a star-forming nebula that is optically thick to photons of the Lyman series but thin to Ly$\mathrm{\alpha}$ and the higher hydrogen series (e.g., Balmer, Paschen, etc), we can obtain the intrinsic ratio between any two hydrogen line within those series. Following \cite{Osterbrock06} and assuming typical of star-forming region conditions (i.e., $\mathrm{n_e\sim100\,cm^{-3}}$ and $\mathrm{T_e\sim10^4\,K}$) we obtain that the intrinsic ratio between the Pa$\mathrm{\beta}$ and H$\mathrm{\alpha}$ lines is $\mathrm{(H\alpha/Pa\beta)_{int} = 17.6}$. Dust grains partly absorb the energy emitted by massive stars in the UV/optical and re-emit such energy as part of the infrared continuum. Thus it is expected that the observed hydrogen ratios differ from the intrinsic value and such difference should act as a proxy for the dust attenuation. In our case, Pa$\mathrm{\beta}$ (12820\AA) is significantly less affected by dust attenuation than H$\mathrm{\alpha}$, as the former lies in the rest-frame NIR. To quantify the amount of nebular dust attenuation present within our sample we follow a similar method to that outlined by previous authors (e.g., \cite{Momcheva13}; \cite{Dominguez13}:
\begin{equation}
\label{EQ:Av}
\mathrm{A_\lambda = \kappa(\lambda)\times E(B-V)}
\end{equation}
where $\mathrm{A_\lambda}$ is the nebular attenuation, $\mathrm{\kappa}$ is the \cite{Calzetti2000} reddening curve, and E(B-V) is the nebular color excess ($\mathrm{E(B-V)=(B-V)_{obs}-(B-V)_{int}}$). The \cite{Calzetti2000} reddening curve provides specific values at the wavelength of the Pa$\mathrm{\beta}$ and H$\mathrm{\alpha}$ emission lines in the form $\mathrm{\kappa(\lambda)=2.659\times(-1.857+1.040/\lambda)+R_V}$ for the wavelength range $\mathrm{0.63\mu m\leq\lambda\leq2.20\mu m}$. Thus, the color excess can be defined as follows:
\begin{equation}
\mathrm{E(B-V) = \frac{E(Pa\beta-H\alpha)}{\kappa(Pa\beta)-\kappa(H\alpha)}}
\end{equation}
However, $\mathrm{E(Pa\beta-H\alpha)}$ can be rewritten in terms of the intrinsic and observed $\mathrm{H\alpha/Pa\beta}$ ratios:
\begin{equation}
\mathrm{E(Pa\beta-H\alpha) = -2.5\times\log\frac{(H\alpha/Pa\beta)_{int}}{(H\alpha/Pa\beta)_{obs}}}
\end{equation}
which yields a simple expression to compute the color excess of our sample based on the observed $\mathrm{Pa\beta}$ and $\mathrm{H\alpha}$ emission line fluxes derived from the dual narrow-band imaging observations presented in this work:  
\begin{equation}
\label{EQ:EBV}
\mathrm{E(B-V) = \frac{-2.5}{\kappa(Pa\beta)-\kappa(H\alpha)}\times\log\frac{(H\alpha/Pa\beta)_{int}}{(H\alpha/Pa\beta)_{obs}}}
\end{equation}
Now, Eq.\,\ref{EQ:Av} can be used to compute the nebular attenuation ($\mathrm{A_\lambda}$) at any wavelength. A usual convention is to study the attenuation over the V band and its evolution concerning other physical quantities, which for the \cite{Calzetti2000} reddening curve takes the form $\mathrm{A_V=4.05\times E(B-V)}$. In the following sections, we will apply this definition whenever we refer to the attenuation of our sample. 

Nevertheless, we caution that the dust extinction estimations computed through hydrogen emission line ratios may vary depending on wavelength distance between any pair of chosen lines, particularly when comparing emission line ratios from the same and different hydrogen series (e.g., \citealt{Dannerbauer05}; \citealt{Groves12}; \citealt{Seille24}). In particular, a luminosity deficit in NIR hydrogen lines such as Br$\mathrm{\gamma}$ and Pa$\mathrm{\alpha}$ has been reported in the past for (U)LIRGs (e.g., \citealt{Goldader95,Goldader97b,Goldader97a}; \citealt{Valdes05}; \citealt{Dannerbauer05}) with respect to the expectations from the bolometric infrared luminosity of these sources. The reason for this deficit is usually linked to starburst activity, which promotes the surge of highly dust-attenuated regions (particularly in central starburst, \citealt{Valdes05}) partly absorbing the light of these emission lines and causing an average $\mathrm{Pa\alpha}$ deficit of 40-60\% in local (U)LIRGs (\citealt{Dannerbauer05}). In the Spiderweb protocluster field, \cite{Dannerbauer14} reported the presence of a dozen SMGs associated with its large-scale structure. These sources are good candidates to host starburst activity given their dusty nature albeit only two of them in this work (ID 73 and 93) are covered by our JWST narrow-band observations. 

\subsection{Environmental proxies}
\label{SS:EnvProxies}

The definition of environment and its influence on the evolution of star-forming disk galaxies in clusters have been well established at $\mathrm{z<1}$ (e.g., \citealt{Boselli22}). This picture becomes less clear at the cosmic noon and beyond, when galaxy evolution proceeds at a faster pace (\citealt{Thomas10}) and the largest and most massive large-scale structures are still in the early stages of their assembly (e.g., \citealt{Muldrew15}; \citealt{Overzier16}; \citealt{Alberts22}). Thus, the absence of virialized massive haloes, the presence of abundant local density peaks, and the heterogenous sampling of overdensities at $\mathrm{z>2}$ introduce severe biases to the environment definition. This work explores three of the most common environmental proxies used in the literature: phase-space diagrams, local density, and clustercentric distance. The phase-space diagram is commonly used to search for halo-mass-related (or large-scale-structure) environmental effects, as it offers the possibility to examine the virialized region of an overdensity through its most basic properties ($\mathrm{M_{200}}$, $\mathrm{R_{200}}$, and $\mathrm{\sigma_{cl}}$). Despite protoclusters not being fully virialized structures, \cite{DiMascolo23} have recently proven the presence of a nascent extended ICM in the Spiderweb protocluster, hinting that this structure might be already en route towards virialization and ensuring the fate of this structure as a future galaxy cluster. In addition and to include both our spectroscopically and narrow-band-only HAEs samples, we use the projected clustercentric distance as an environmental proxy by taking the Spiderweb radio galaxy as the center of this structure. However, protoclusters are often made up of a collection of infalling groups representing local density peaks where gravitational interactions may be more frequent. This situation has been observed within higher redshift protocores (e.g., \citealt{Long20}; \citealt{Calvi21}; \citealt{Hill22}; \citealt{Zhou24}) which are usually populated by dusty starbursts. Thus, we also trace the environment of our sources by measuring their local density defined as the projected distance enclosing up to 5 neighboring galaxies (i.e., $\mathrm{\Sigma_5}$). To this end, we collect a sample of 147 unique sources in the Spiderweb protocluster encompassing all known galaxy populations in this field (Spiderweb complex from \citealt{Kuiper11}; LAEs from \citealt{Pentericci00}; SMGs from \citealt{Dannerbauer14}; CO emitters from \citealt{Emonts18}, \citealt{Tadaki19}, and \citealt{Jin21}; HAEs from \citealt{Koyama13}, \citealt{Shimakawa18b}, and \citealt{PerezMartinez23}; and the newly selected PBEs from S24) and compute $\mathrm{\Sigma_5}$ through the following expression: 
\begin{equation}
   \Sigma_N=\frac{N}{\pi R^2_{N-1}}
\label{LocalDensity}
\end{equation}
where $\mathrm{R_{N-1}}$ is the distance to the $\mathrm{n^{th}-1}$ neighbor of a given object. $\mathrm{\Sigma_5}$ represents a good compromise between measuring small but representative groups of galaxies and excluding the smallest density peaks caused by doublets or triplets. None of these environmental proxies are absent of biases due to projection effects or heterogeneous sampling across the surveyed field. However, the combined use of these three approaches would help to constraint the environment of our sample as shown by previous protocluster studies (e.g., \citealt{Shimakawa18b}, \citealt{Wang18}; \citealt{PerezMartinez23,PerezMartinez24}; \citealt{Toshikawa24}).

\begin{figure*}
\centering
\includegraphics[width=0.5\linewidth]{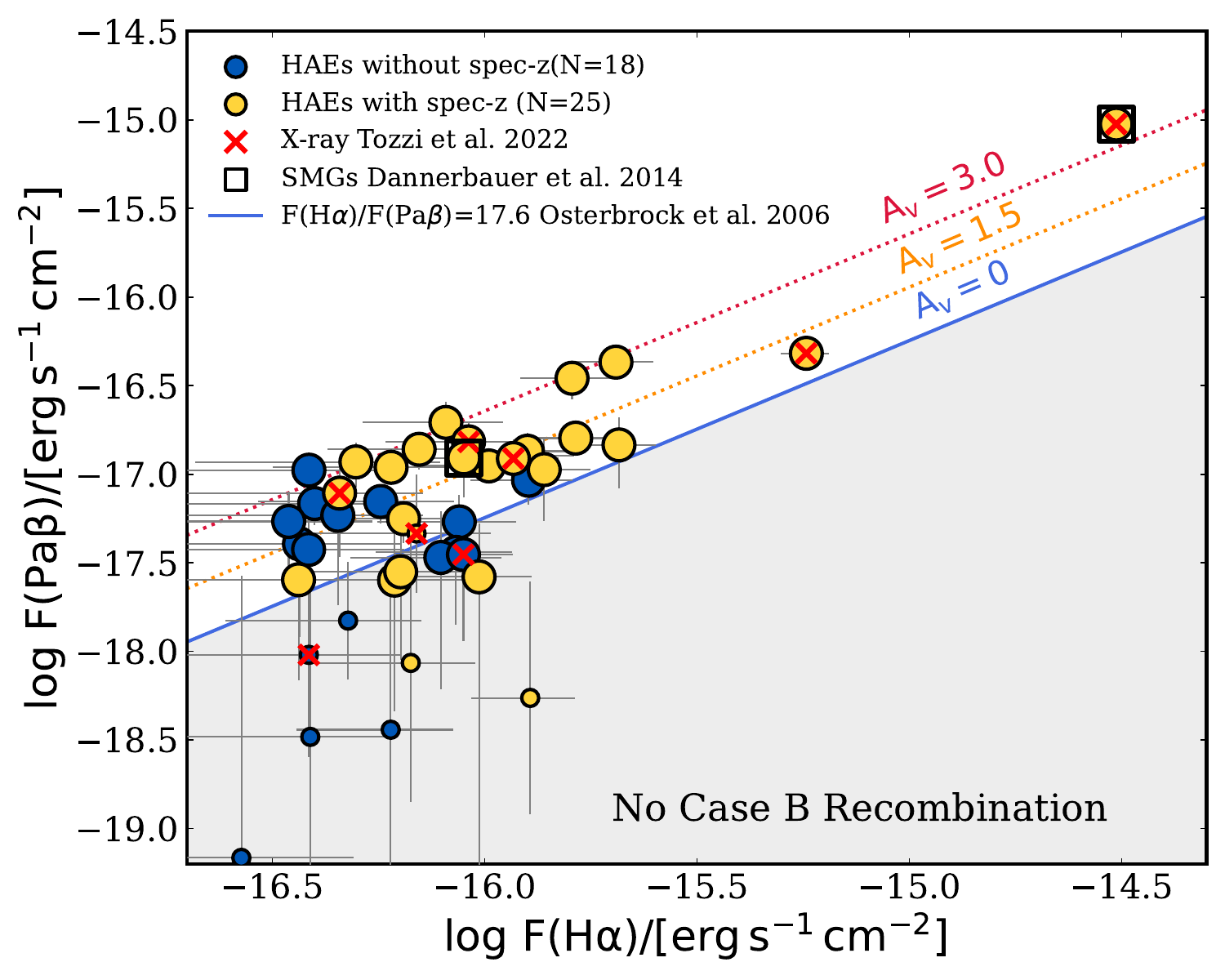}~
\includegraphics[width=0.5\linewidth]{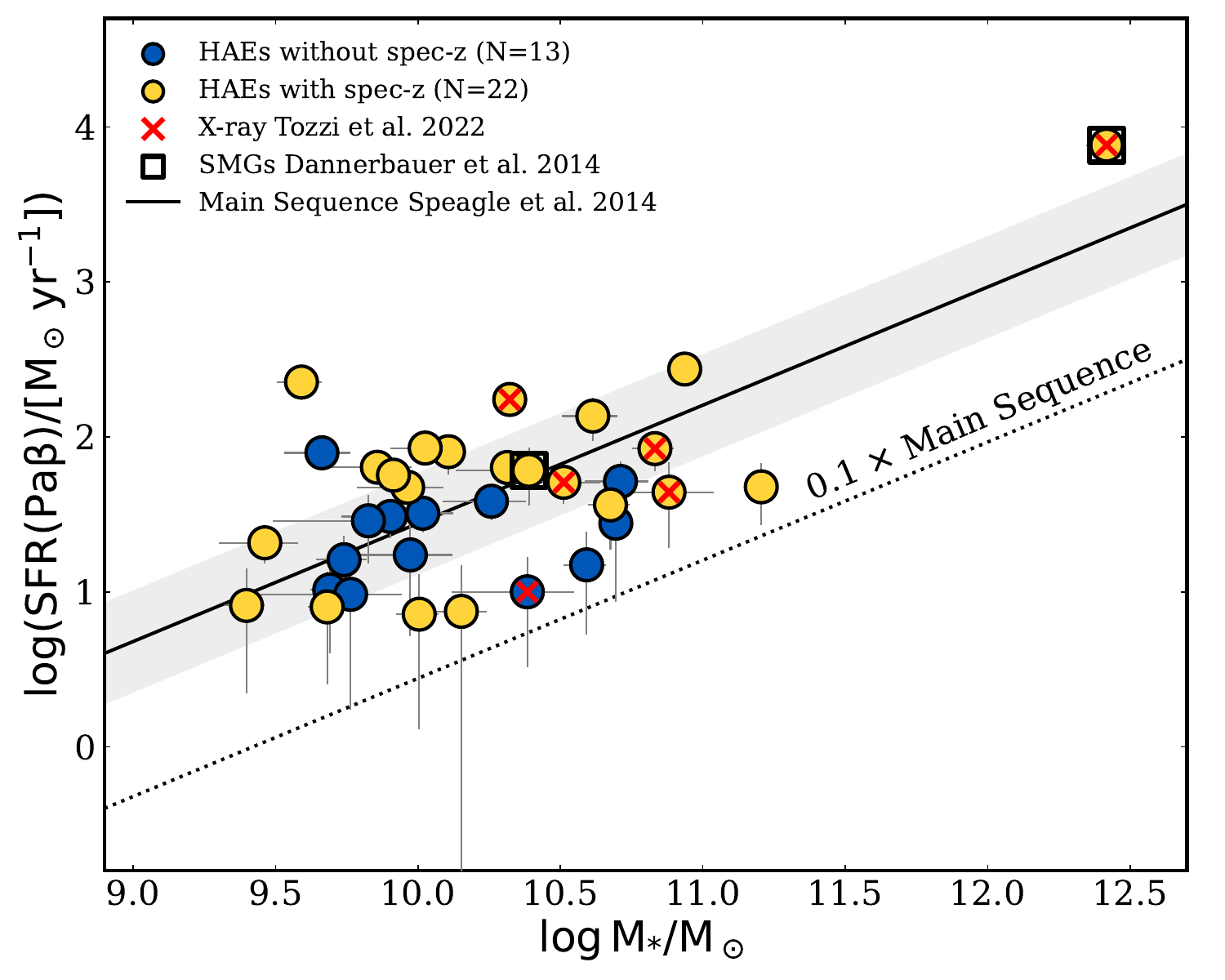}\par
\caption{Left: H$\mathrm{\alpha}$ vs Pa$\mathrm{\beta}$ flux ratio diagram for HAEs in the Spiderweb protocluster. Yellow and blue circles show the emission line fluxes of our sample of HAEs with and without spectroscopic redshift respectively. The solid and dotted lines represent increasing values of nebular extinction $\mathrm{A_V}$ (see Sect.\,\ref{SS:DustExt}). The grey shaded area displays the forbidden region where our assumptions of Case B recombination are no longer valid. Right: Main sequence diagram. Symbols and colors of the HAE samples follow the same scheme shown in the left panel. The solid line and the grey area display the location of the Main Sequence of star formation from \cite{Speagle14}. The dotted line marks the boundary of 1\,dex below the main sequence.}
\label{F:Ratios}
\end{figure*}

\section{Results and Discussion}
\label{S:Results}

\subsection{Dust attenuation and star-formation properties}
\label{SS:DustProp}

Our first goal is to investigate the general properties of the H$\mathrm{\alpha}$ and Pa$\mathrm{\beta}$ emission line flux within our sample and compute the nebular dust attenuation of our sources following the procedure outlined in Sect.\,\ref{SS:DustExt}. The left panel of Fig.\,\ref{F:Ratios} displays the distribution of our narrow-band only (blue) and spectroscopic (yellow) HAE samples in terms of their emission line fluxes. Most HAEs within the Spiderweb protocluster display V-band dust attenuation ranging at $\mathrm{A_V=0-3}$ mag which is equivalent to $\mathrm{A_{H\alpha}=0-2.5}$ mag. These values are typical for star-forming galaxies at the cosmic noon (e.g., \citealt{Pannella15}; \citealt{Wisnioski19}), in contrast with the typical attenuation values from coeval SMGs for which $\mathrm{A_{V}\gtrsim3}$ (e.g., \citealt{Smail23}; \citealt{Polletta24}; \citealt{Gillman24}).

Some sources lie below $\mathrm{A_V=0}$ mag (solid line) in Fig.\,\ref{F:Ratios}, which marks the limit for no attenuation assuming the case B recombination for $\mathrm{n_e\sim100\,cm^{-3}}$ and $\mathrm{T_e\sim10^4\,K}$ (\citealt{Osterbrock06}). Possible variations in electron temperature or density within the Case B recombination assumption can not explain these offsets, as they would only account for up to 0.06 magnitudes in the most extreme case (i.e., $\mathrm{n_e\sim1000\,cm^{-3}}$ and $\mathrm{T_e\sim2\times10^4\,K}$). We identify 8 HAEs whose uncertainties in Pa$\mathrm{\beta}$ are greater than the measured flux value (i.e., $\mathrm{S/N(PaB)<1}$, small yellow and blue circles in Fig.\,\ref{F:Ratios}). Interestingly, 7 of them lie well inside the forbidden Case B recombination region in this figure, thus suggesting that their location may be the product of the uncertainties in the measurements rather than due to special conditions in their ISM. Furthermore, these 8 sources display very compact apparent sizes in the JWST/NIRCam images, similar to the 13 sources that were previously removed from our sample in Sect.\,\ref{SS:EL}) due to their lack of $\mathrm{Pa\beta}$ emission. Thus, we also remove these sources from further analysis given their unreliable nature. As a consequence, our final sample is composed of 22 HAEs with spectroscopic redshift and 13 HAEs without. Finally, 6 sources in our final sample still lie within the forbidden case B region (larger blue and yellow circles) but close to the boundary $\mathrm{A_V=0}$, making them compatible with the absence of dust within the uncertainties of our measurements.

We note that the metallicity correction carried out in Sect.\,\ref{SS:EL} intends to remove the [N{\sc{ii}}] contribution from the narrow-band flux based on the average mass-metallicity relation of \cite{PerezMartinez23} in this protocluster. However, the scatter of this relation is significant, and moderate variations of metallicity over the average value may naturally shift some of these sources inside the permitted Case B region. It is for this reason that hereafter we assign a minimum value of $\mathrm{A_V=0}$ mag (see also Table\,\ref{T:appendix}) to these few sources lying slightly below such boundary. In addition, we crossmatch our HAE samples with the existing X-ray emitting sources (i.e., likely AGN) studied by \cite{Tozzi22a}. We find 8 sources overlap with our sample including the Spiderweb galaxy at the massive stellar mass end. The distribution of these sources in Fig.\,\ref{F:Ratios} do not correlate with dust extinction with three sources lying at $\mathrm{A_V>1.5}$, another 3 objects displaying $\mathrm{0<A_V<1.5}$, and two objects residing within the case B forbidden region, with one of them being previously discarded due to its poor S/N. Finally, we also inspect the population of SMGs reported by \cite{Dannerbauer14} in this field. Only four (DKB02, DKB07, DKB13, and DKB15) out of their sixteen sources are covered by this JWST/NIRCam program. However, DKB13 was already labeled as a non-member by \cite{Tanaka10}. Of the remaining three sources, only two of them (DKB07 and DKB15) have an HAE counterpart (\citealt{Shimakawa18b}). DKB07 is identified as a bright submillimeter source overlapping with the Spiderweb complex including its radio galaxy, which shows $\mathrm{A_V=3.6}$ mag based on its measured $\mathrm{H\alpha/Pa\beta}$ ratio. On the other hand, DKB02 and DKB15 were only classified as a tentative member in \citealt{Dannerbauer14}, with the latter one being spectroscopically confirmed ($\mathrm{z=2.152}$) by \citealt{PerezMartinez23}) and displaying an extinction value of $\mathrm{A_V=2.2}$ mag in this work. This shows the potential of rest-frame NIR emission lines such as Pa$\mathrm{\beta}$ to confirm the membership of submillimetre-identified sources. Finally, both DKB07 and DKB15 display below average dust extinction values compared to the coeval field SMGs reported in recent works albeit both lie well inside the expected scatter for this quantity (e.g., \citealt{Dudzeviciute20}; \citealt{Polletta24}; \citealt{Gillman24}).

Having explored the dust attenuation in the left panel of Fig.\,\ref{F:Ratios}, we examine the star formation activities (SFR) of our sample combining the measured NIR Pa$\mathrm{\beta}$ emission line fluxes and the extinction values computed through the hydrogen decrements. The reason to choose Pa$\mathrm{\beta}$ over H$\mathrm{\alpha}$ is that the former is isolated within the narrow-band filter bandwidth, while H$\mathrm{\alpha}$ needs to be corrected from {\sc{[nii]}} contamination (see Sect.\,\ref{SS:EL}). However, both hydrogen lines can be affected by the presence of AGN ionization. The right panel of Fig.\,\ref{F:Ratios} displays the SFR($\mathrm{Pa\beta}$) of our sources as a function of the stellar mass values reported in \cite{Shimakawa24}. We applied the star formation calibration by \cite{Kennicutt98} renormalized for a \cite{Chabrier03} IMF and assuming the intrinsic ratio $\mathrm{H\alpha/Pa\beta=17.6}$ from the case B recombination. In addition, we apply the nebular extinction computed above to the observed Pa$\mathrm{\beta}$ emission line flux following $\mathrm{A_{Pa\beta}\approx0.3\,A_V}$ for a \cite{Calzetti2000} extinction law. The diagram shows a mix of behaviors albeit most sources are compatible with the expectations from the Main Sequence of star formation by \cite{Speagle14} at $\mathrm{z=2.16}$. These results are qualitatively in agreement with those of \citealt{PerezMartinez23}, who used VLT/KMOS to compute H$\mathrm{\alpha}$-based SFR with dust extinction from SED fitting. Thus, our sample of HAEs resembles typical cosmic noon star-forming galaxies both in terms of dust attenuation and star formation. Nevertheless, we note that attenuation values computed through UV and optical tracers tend to be biased towards the less extinct areas of a given object, which would result in the underestimation of the average attenuation value. In addition, we acknowledge that our sample selection (HAEs, see Sect.\,\ref{S:OBs}), flux limits, and emission line ratio requirements (Sect.\,\ref{SS:EL}) exclude from our analysis a population of objects that either have low intrinsic H$\alpha$ flux but could in some cases display moderate amounts of dust (e.g., low-mass galaxies) or objects that are extremely dusty by nature (e.g., bright SMGs), as they would likely not enter into the HAE class defined by \cite{Shimakawa24} due to their attenuation. Some of these sources have been primarily detected at submillimetre wavelengths in past works (e.g., \citealt{Dannerbauer14}; \citealt{Jin21}; \citealt{ChenZ24}), but also recently in S24 as PBEs lacking H$\mathrm{\alpha}$ emission, and require future follow-up observing programs to establish their physical properties.

\subsection{Dust attenuation and stellar mass}
\label{SS:DustProp2}

\begin{figure}    
\centering
\includegraphics[width=\linewidth]{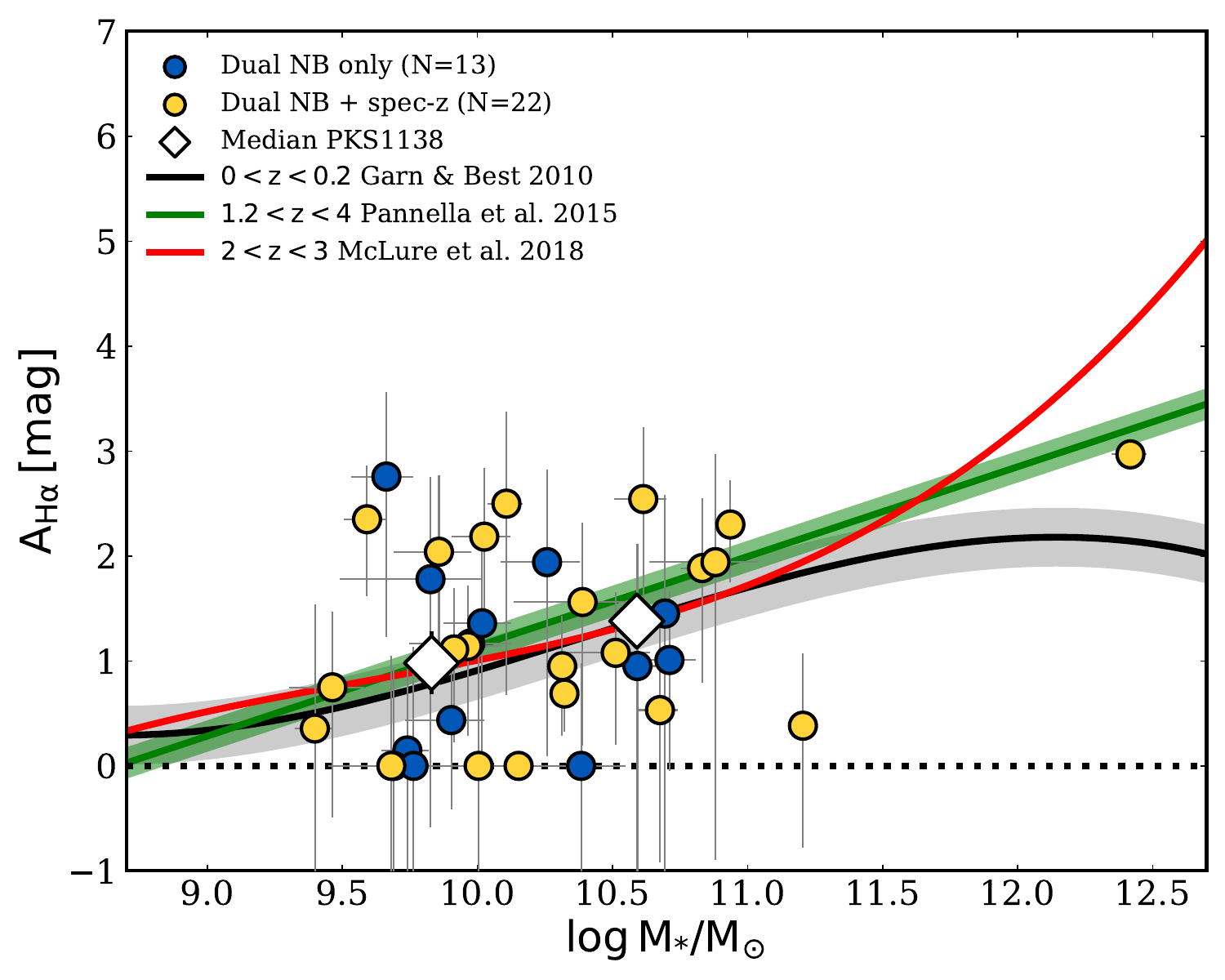}\par
\caption{Dust attenuation over H$\mathrm{\alpha}$ vs stellar mass diagram. Symbols and colors from our sample follow the same scheme as in previous figures. Big white diamond display the median values for our sample within the stellar mass bins $\mathrm{9.4<\log M_*/M_\odot<10.2}$ and $\mathrm{10.2<\log M_*/M_\odot<11.0}$. Solid lines display the dust attenuation laws from \cite{Garn10}, \cite{Pannella15}, and \citealt{McLure18} renormalized for the \cite{Chabrier03} IMF and the H$\mathrm{\alpha}$ wavelength following a \citealt{Calzetti2000} extinction law. The green shaded area depicts the 0.3 mag evolution for \cite{Pannella15} relation within their redshift range.}
\label{F:Dustmass}
\end{figure}

Previous works have reported a correlation between dust extinction and stellar mass (e.g., \citealt{Garn10}) since dust represents the byproduct of star formation, and the stellar mass embodies the cumulative star formation history of a given object. Thus, we examine the link between these two quantities within our sample in Fig.\,\ref{F:Dustmass}. Following the convention of \cite{Garn10} we  display our values for $\mathrm{A_{H\alpha}}$ as a function of stellar mass after applying $\mathrm{A_{H\alpha}=0.82\,A_V}$ following the \citealt{Calzetti2000} extinction law. The scatter of our sample is too large to draw conclusions from their distribution with no clear correlation present within our data. We split our sample into two stellar mass bins given by $\mathrm{9.4<\log M_*/M_\odot<10.2}$ (18 sources) and $\mathrm{10.2<\log M_*/M_\odot<11.0}$ (15 sources) and compute their median values to search for any trend. In the process, we excluded the Spiderweb galaxy and another object from this analysis as both of them lie beyond the proposed stellar mass bins. Our binned results (white diamonds) display a mild increase of $\mathrm{A_{H\alpha}}$ with stellar mass which resembles the predictions made by \cite{Garn10} in the local universe. By comparison, we display the relations the $\mathrm{1.2<z<4}$ sample of \cite{Pannella15} and the $\mathrm{2<z<3}$ work by \cite{McLure18}. These two latter works extract their dust attenuation values from the UV continuum of their sources, and thus their stellar mass vs dust extinction relations had to be shifted to the wavelength of $\mathrm{H\alpha}$ and converted into nebular attenuation as it is the case of our work and \cite{Garn10}. We follow the same method outlined in \cite{McLure18} for this purpose. Finally, we convert the stellar masses by \cite{Pannella15} from the \cite{Salpeter55} IMF to \citealt{Chabrier03}. These three relations overlap with each other within the scatter and also match our binned results for the examined mass range ($\mathrm{9.4<\log M_*/M_\odot<11.0}$). This result reinforces previous findings by \cite{PerezMartinez23,PerezMartinez24} showing that most HAEs in this protocluster behave as typical star-forming galaxies at the cosmic noon, although with a significant scatter in their properties. This may reflect the impact of secondary effects derived from their overdense environment and the intrinsic stochastic nature of the star formation histories of different objects.

\subsection{Environmental analysis}
\label{SS:EnvResults}

\begin{figure}    
\centering
\includegraphics[width=\linewidth]{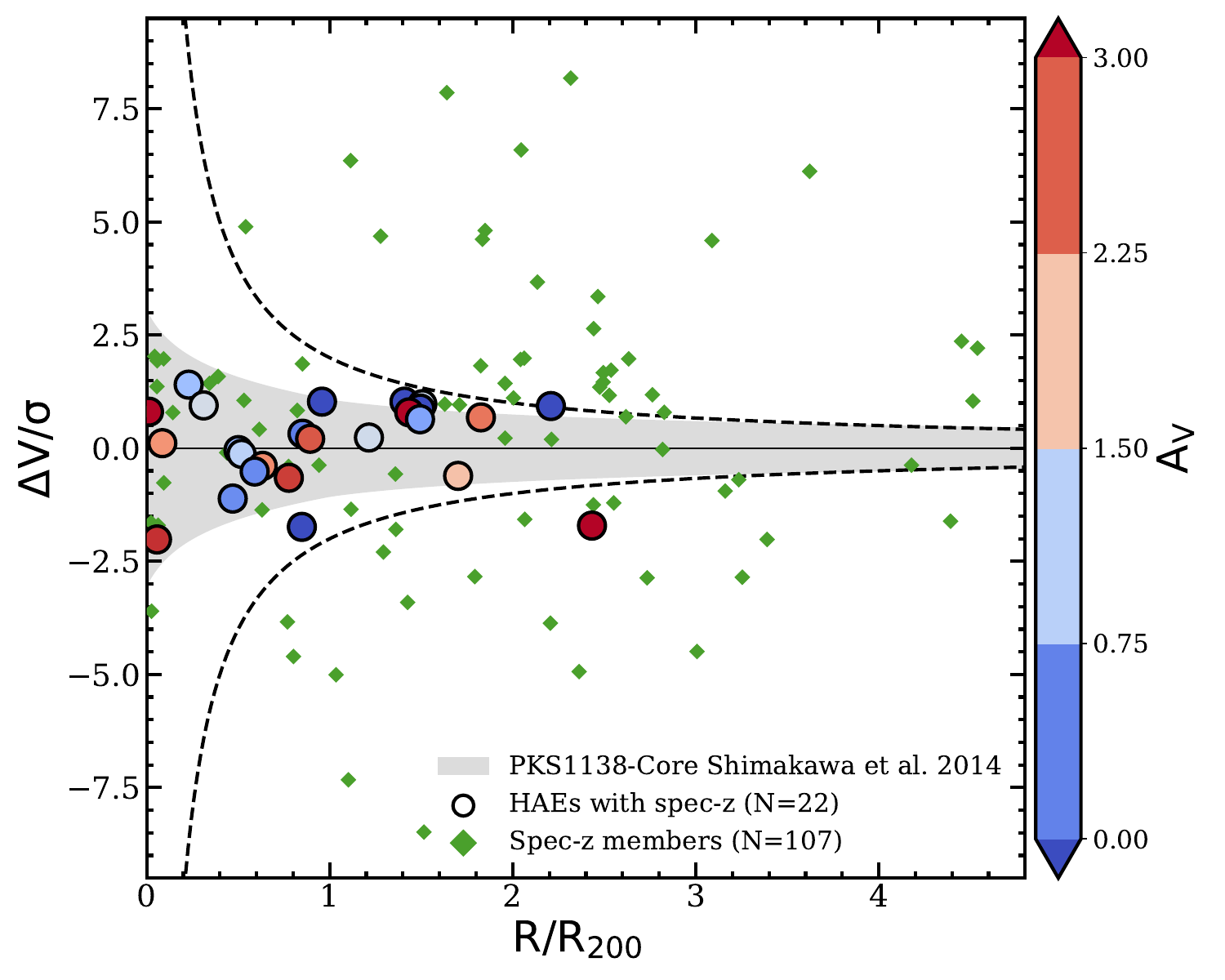}\par
\caption{Phase-space diagram. Our spectroscopic sample of HAEs is displayed by circles color-coded according to the nebular attenuation $\mathrm{A_V}$. The green diamonds display additional spectroscopically confirmed protocluster members from different sources (e.g., \citealt{Pentericci00}; \citealt{Croft05}; \citealt{Kuiper11}; \citealt{Shimakawa15}; \citealt{Emonts18}; \citealt{Tadaki19}; \citealt{Jin21}; \citealt{PerezMartinez23}). The shaded grey area represent the virialized region according to the values for $\mathrm{M_{200}}$, $\mathrm{R_{200}}$, and $\mathrm{\sigma_{cl}}$ derived in \citealt{Shimakawa14}. The dashed line marks the boundary between the accreted and outskirts regions ($\mathrm{\eta=(R_{\mathrm{proj}}/R_{200})\times(\left | \Delta v \right |/\sigma_{cl})=2}$) of an overdensity as defined by \citealt{Noble13}.}
\label{F:PhaseSpace}
\end{figure}

Dust production is a byproduct of star formation activities and as such, it is related to their main drivers. Protoclusters are thought to be exceptional sites of star formation due to enhanced gas accretion from the cosmic web (e.g., \citealt{Chiang17}, \citealt{Umehata19}; \citealt{Daddi22a}; \citealt{Remus23}) as well as due to the higher number of gravitational interactions between gas-rich galaxies in overdense environments (e.g., \citealt{Lotz13}; \citealt{Hine16}; \citealt{Watson19}; \citealt{LiuS23}; \citealt{Naufal23}; \citealt{Shibuya24}). On the other hand, environmental effects such as ram pressure stripping or strangulation in the presence of an intracluster medium (ICM) act in the opposite direction, albeit these effects are frequent only within matured clusters typically found at later epochs. Nevertheless, it is reasonable to expect that dust production and by extension star formation correlate with some local or large-scale environmental proxy within the Spiderweb protocluster. We test this hypothesis in Fig.\,\ref{F:PhaseSpace} where we display the phase-space diagram of the Spiderweb protocluster assuming the values for $\mathrm{M_{200}}$, $\mathrm{R_{200}}$, and $\mathrm{\sigma_{cl}}$ derived in \cite{Shimakawa14}. The grey shaded area in this diagram depicts the virialized region based on those values, while the dashed line marks the boundary between the accreted and outskirts regions ($\mathrm{\eta=(R_{\mathrm{proj}}/R_{200})\times(\left | \Delta v \right |/\sigma_{cl}=2}$) of an overdensity as defined by \cite{Noble13} (see also \citealt{Haines12}). We restrict our primary sample to those objects with spectroscopic redshift as this is a necessary condition to determine their velocity offset along the line of sight with respect to the reference redshift of the protocluster ($\mathrm{z=2.156}$). Our sample was color-coded according to the values of nebular extinction ($\mathrm{A_V}$) measured following Sect.\,\ref{SS:DustExt}. We find that the majority of our spectroscopic HAEs reside within the virialized region. This is expected due to their narrow-band selection (\citealt{Koyama13}; \citealt{Shimakawa18b}; \citealt{Daikuhara24}) and their proximity to the defined center of the protocluster (i.e., the Spiderweb Galaxy) as shown in Fig.\,\ref{F:Map}. By comparison, other spectroscopic samples (e.g., CO emitters from \citealt{Jin21}) trace the same overdensity albeit on a much larger scale across the line of sight. Nevertheless, we find no correlation between the position of our galaxies across the protocluster large-scale structure and their dust attenuation levels. We repeat this analysis by examining the pseudo-3D distribution of our targets. This time we convert the difference between their redshifts and the systemic redshift of the Spiderweb radio galaxy to a third spatial dimension along the line of sight. However, our results yield no correlation between the spatial distribution of our spectroscopically confirmed HAEs and their extinction values. These findings qualitatively supports the results of \cite{Dannerbauer14}, who found that most of the SMGs within this protocluster avoid its central regions and spread across the outer regions of the protocluster, suggesting an environment-independent origin (see also \citealt{Smail14,Smail24}; \citealt{Zhang22}). 
\begin{figure*}
\centering
\includegraphics[width=0.5\linewidth]{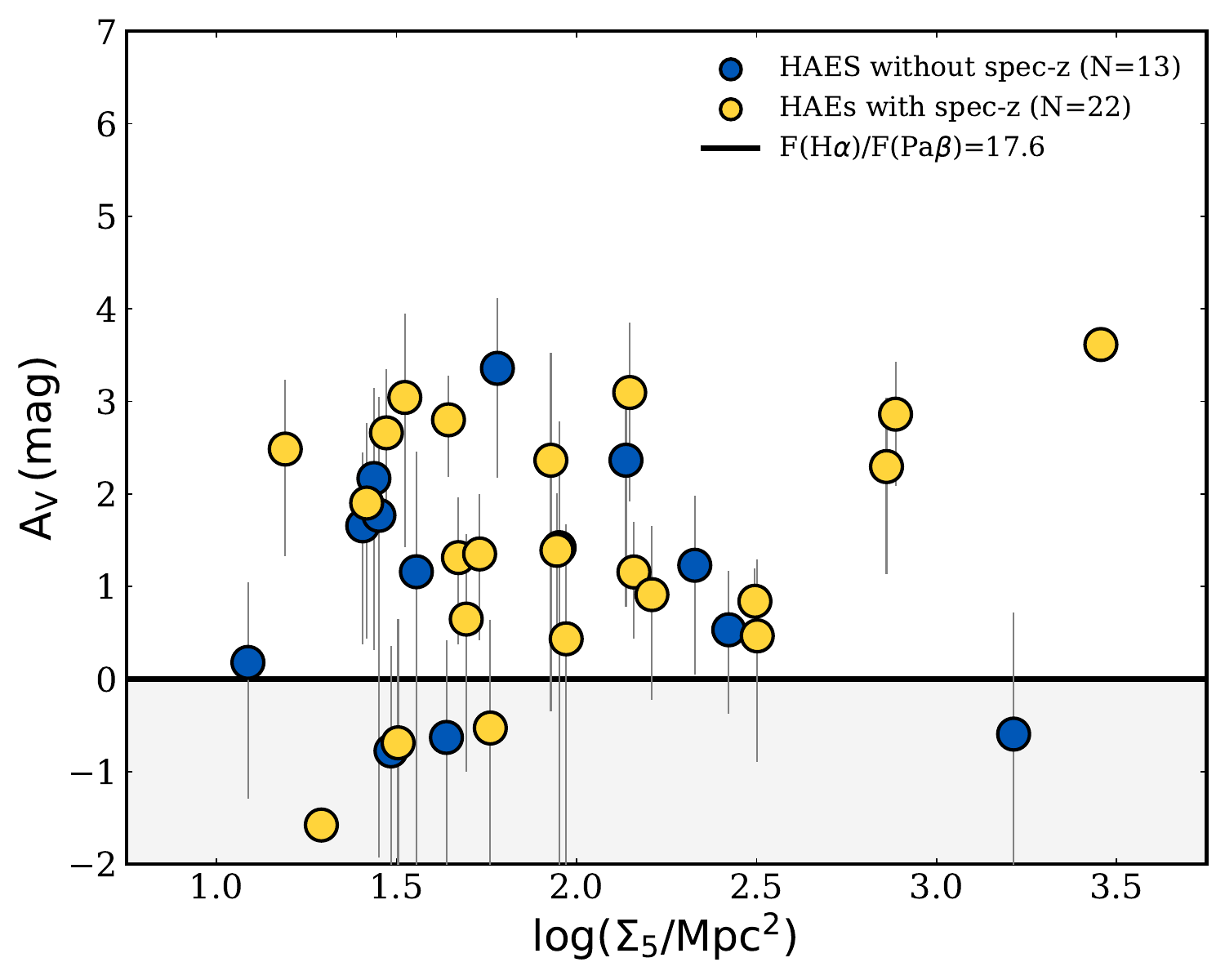}~
\includegraphics[width=0.5\linewidth]{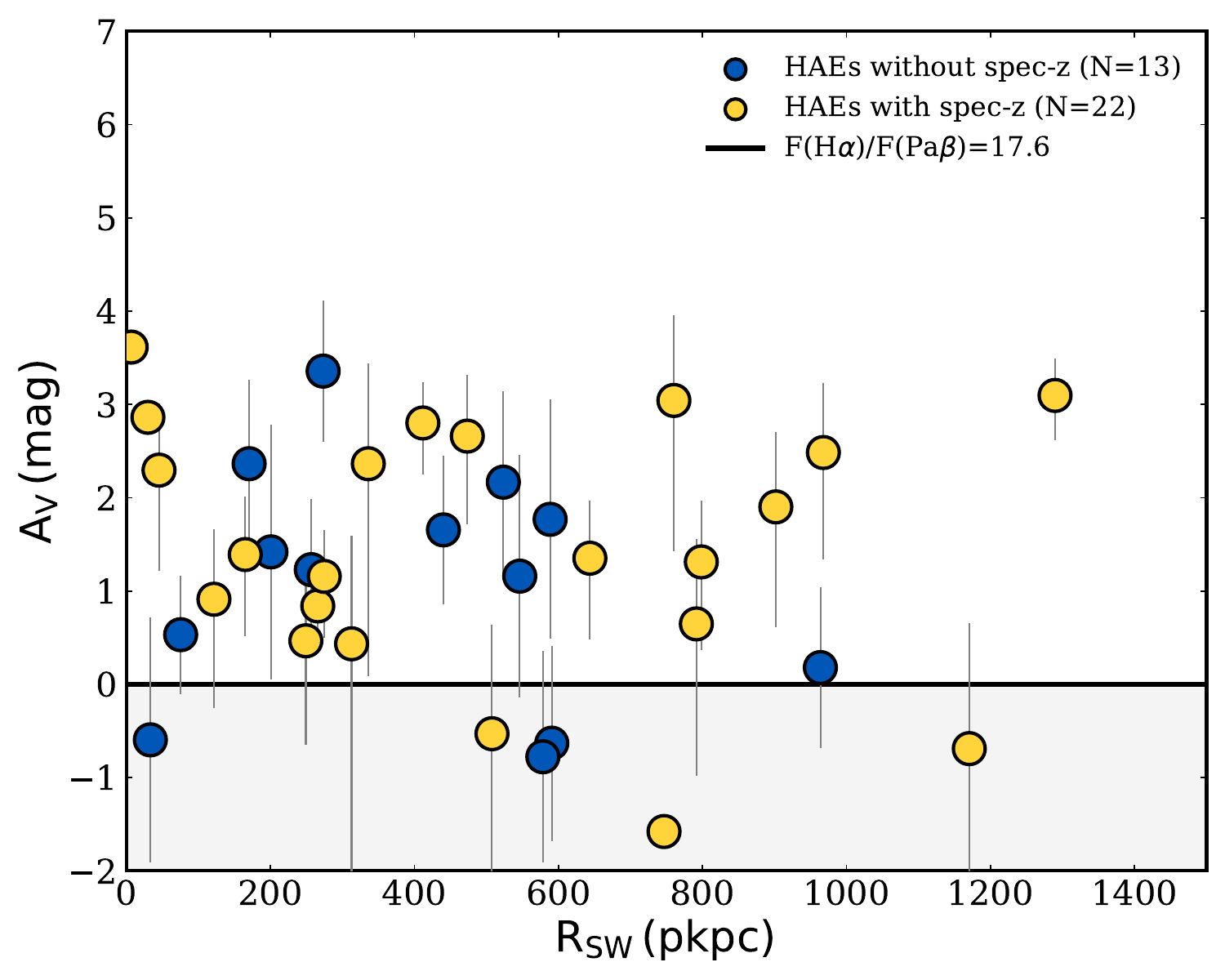}~
\caption{Nebular dust attenuation ($\mathrm{A_V}$) as a function of local density ($\mathrm{\Sigma_5}$) taking as a reference five neighboring galaxies (left), and the projected clustercentric distance towards the Spiderweb (SW) galaxy. Yellow and blue circles represent our samples of HAEs with and without spectroscopic redshif as in previous figures. We allow for negative values $\mathrm{A_V}$ (grey shaded areas) to display those objects whose H$\mathrm{\alpha}$/Pa$\mathrm{\beta}$ ratios are higher than 17.6 and thus not compatible with the Case B recombination (see Sect.\,\ref{SS:DustProp}).}
\label{F:S5}
\end{figure*}

We propose three possible explanations for this result: First, the higher levels of dust extinction found in some objects may be the result of past but recent violent episodes of star formation triggered by gravitational interactions such as mergers or close encounters in local density peaks distributed across the protocluster structure. Second, dust extinction may be driven by secular star formation processes fed by smooth gas accretion toward the protocluster large-scale structure. Third, sampling bias may be a relevant factor in washing out the signal of any underlying environmental correlation. We test the first possibility in Fig.\,\ref{F:S5} by showing the dust attenuation of our spectroscopic and narrow-band-only samples as a function of local density and clustercentric radius. We see in both cases no clear correlation across the parameter space surveyed, with dusty and dustless galaxies populating the whole local density and clustercentric distance regimes. This result is also consistent with the lack of correlation of H$\mathrm{\alpha}$-based SFR with the environment described by \cite{PerezMartinez23} in this same protocluster. We note that in the densest region ($\mathrm{\log\Sigma_5>2.5\,Mpc^2}$) three out of the four surveyed galaxies display relatively high levels of nebular attenuation ($\mathrm{A_V>2}$). Still, one of these objects is the Spiderweb galaxy whose nature is special as the proto-BCG of this protocluster and the two others are massive galaxies for which higher levels of dust extinction are expected. These results suggest that local density peaks or closeness to the center of the protocluster are not promoting dust production within our HAEs, implying that environmental effects related to a higher number density of objects (e.g., gravitational interactions) may not be playing a relevant role at the moment in driving the star formation and dust production within our sample. We note that the Spiderweb protocluster is the only overdensity at $\mathrm{z>2}$ with unambiguous signs of a nascent ICM (\citealt{DiMascolo23}), thus confirming its nature as transitional large-scale structure or in other words a bonafide galaxy cluster in formation. In such a scenario, the bursty star formation activities reported in high redshift protocores dominated by interacting SMGs (\citealt{Oteo18}; \citealt{Long20}; \citealt{Hill22}; \citealt{Calvi23}; \citealt{Zhou24}) would have ceased several hundred Myrs ago, yielding the so-called Spiderweb complex (\citealt{Miley06}; \citealt{Kuiper11}) and a few of the surrounding massive HAEs. If these objects would have resided in an underdense environment, their star-formation could have been rapidly quenched after the burst due to insufficient accretion (e.g., \citealt{Kimmig24}). However, the highly overdense region that surrounds them promotes the infall of cold gas through filaments reaching the center of the halo and supporting relatively normal levels of star formation that we detect for the time being (e.g., \citealt{Remus24}). Similarly, most low-mass galaxies across the protocluster structure would be able to proceed with their galaxy build-up as shown by \cite{Daikuhara24} for the time being and until the increasing temperature and density of the ICM promote the onset of classical environmental effects such as ram pressure stripping or starvation at later epochs (see \citealt{Boselli22} for a review). On the other hand, the current population of SMGs lying at the outskirts of the protocluster (\citealt{Dannerbauer14}) are likely newcomers whose infall happened recently and thus, their observed properties would have been determined independently of their current environment. These objects, which might be enduring starburst episodes may not have the opportunity to rekindle their star formation as the Spiderweb protocluster progresses in their virialization, and thus they are good candidates to form the massive end of the future red sequence by $\mathrm{z=1}$ (e.g., \citealt{Ivison13}; \citealt{Smail14}).

Regarding the third possibility, we admit that our final sample counts with a relatively low number of objects (N=35) compared with the overall population of the protocluster ($N>100$). However, we can adequately cover the mass range $\mathrm{9.4<\log M_*/M_\odot<11}$ for HAEs within the central volume of the protocluster finding no relevant environmental dependence, as in previous studies in this field. We acknowledge, however, that scaling relations involving the dust attenuation tend to have relatively large scatter (e.g., \citealt{Garn10}; \citealt{Groves12}; \citealt{Dominguez13}; \citealt{Kashino13}; \citealt{Pannella15}), and thus, higher number statistics may unveil mild trends that remain hidden to our current understanding.

\section{Conclusions}
\label{S:Conclusions}

The advent of JWST has opened a new window to use rest-frame NIR dust and star formation indicators at $\mathrm{z>2}$. In this work, we have used new JWST/NIRCam PaB narrow-band imaging to investigate the dust attenuation and star formation activities of a sample of HAEs in the Spiderweb protocluster at $\mathrm{z=2.16}$. Combined with ancillary Subaru/MOIRCS H$\mathrm{\alpha}$ narrow-band observations, we have obtained reliable H$\mathrm{\alpha}$/Pa$\mathrm{\beta}$ ratios for 35 sources, and analyze their internal properties with respect to several environmental indicators. Our conclusions can be summarized as follows:
\begin{enumerate}
\item Most HAEs in the Spiderweb protocluster display H$\mathrm{\alpha}$/Pa$\mathrm{\beta}$ ratios yielding nebular dust attenuation of the order of $\mathrm{Av=0-3}$ magnitudes assuming hydrogen emission line ratios dominated by the case B recombination ($\mathrm{T_e=10^4\,K}$ and $\mathrm{n_e=10^2\, cm^{-3}}$, \citealt{Osterbrock06}) and the \cite{Calzetti2000} extinction law. These values are within the range expected for field galaxies at the cosmic noon (e.g., \citealt{Pannella15}; \citealt{McLure18}).
\item A quarter of our original HAE sample displays H$\mathrm{\alpha}$/Pa$\mathrm{\beta}$ ratios not compatible with the case B recombination (i.e., below $\mathrm{A_V=0}$). The H$\mathrm{\alpha}$ and Pa$\mathrm{\beta}$ flux uncertainty of these objects tend to be similar to their respective flux measurements. This would suggest that the ISM conditions within these objects may not be different than those of the rest of our sample, albeit deeper observations would be required to confirm such a scenario.
\item Applying these new $\mathrm{Pa\beta}$ based dust attenuation values to our sample of HAEs, we find that most of them scatter around the main sequence of star formation (\citealt{Speagle14}) thus displaying similar properties to their coeval field counterparts. These results confirm the findings of previous narrow-band and spectroscopic works for HAEs in this protocluster (e.g., \citealt{Shimakawa18b}; \citealt{PerezMartinez23}), and suggest that while HAEs are the most numerous population they are currently evolving in a secular way with no clear evidence of environmental effects promoting or surprising their star formation activities. 
\item We find no correlation between the dustyness of our sample of HAEs and their distribution in phase space (spectroscopic sample), or as a function of the projected clustercentric radius or local density ($\mathrm{\Sigma_5}$). This suggests that dust production or star formation activities may be fueled by relatively smooth cold gas accretion towards these objects while gravitational interactions such as fly-byes and mergers as a consequence of clustering may play a subdominant role for most of them at this stage.
\end{enumerate}

\begin{acknowledgements}

We thank the anonymous referee for his/her constructive feedback, which has contributed to improving this manuscript. This research is based on observations made with the NASA/ESA James Webb Space Telescope obtained from the Space Telescope Science Institute, which is operated by the Association of Universities for Research in Astronomy, Inc., under NASA contract NAS 5–26555. These observations are associated with the JWST/NIRCam program ID\#1572, PI: H. Dannerbauer. NIRCam was built by a team at the University of Arizona (UofA) and Lockheed Martin's Advanced Technology Center, led by Prof. Marcia Rieke at UoA.  This research is based in part on data collected at the Subaru Telescope, which is operated by the National Astronomical Observatory of Japan (NAOJ). We are honored and grateful for the opportunity of observing the Universe from Maunakea, which has cultural, historical, and natural significance in Hawaii. This research made use of Astropy,\footnote{http://www.astropy.org} a community-developed core Python package for Astronomy \citep{Astropy13, Astropy18}. JMPM acknowledges funding from the European Union’s Horizon-Europe research and innovation programme under the Marie Skłodowska-Curie grant agreement No 101106626. HD, YZ, and JMPM acknowledge financial support from the Agencia Estatal de Investigación del Ministerio de Ciencia e Innovación (AEI-MCINN) under grant (La evolución de los c\'umulos de galaxias desde el amanecer hasta el mediod\'ia c\'osmico) with reference (PID2019-105776GB-I00/DOI:10.13039/501100011033). HD, YZ, and JMPM acknowledge financial support from the Ministerio de Ciencia, Innovaci\'on y Universidades (MCIU/AEI) under grant (Construcci\'on de c\' umulos de galaxias en formaci\'on a trav\'es de la formaci\'on estelar oscurecida por el polvo) and the European Regional Development Fund (ERDF) with reference (PID2022-143243NB-I00/DOI:10.13039/501100011033). YK, RS, and TK acknowledge support from JSPS KAKENHI Grant Number 23H01219. PGP-G acknowledges support from grant PID2022-139567NB-I00 funded by Spanish Ministerio de Ciencia, Innovaci\'on y Universidades MCIU/AEI/10.13039/501100011033, FEDER {\it Una manera de hacer Europa}. TK acknowledges financial support from JSPS KAKENHI Grant Numbers 24H00002 (Specially Promoted Research by T. Kodama et al.) and 22K21349 (International Leading Research by S. Miyazaki et al.). YZ acknowledges the support from the China Scholarship Council (202206340048), and the National Science Foundation of Jiangsu Province (BK20231106). C.D.E. acknowledges funding from the MCIN/AEI (Spain) and the "NextGenerationEU"/PRTR (European Union) through the Juan de la Cierva-Formación program (FJC2021-047307-I).
\end{acknowledgements}



\appendix
\restartappendixnumbering
\section{Gallery}
\label{A:Gallery}
\begin{figure*}
\includegraphics[width=0.5\linewidth]{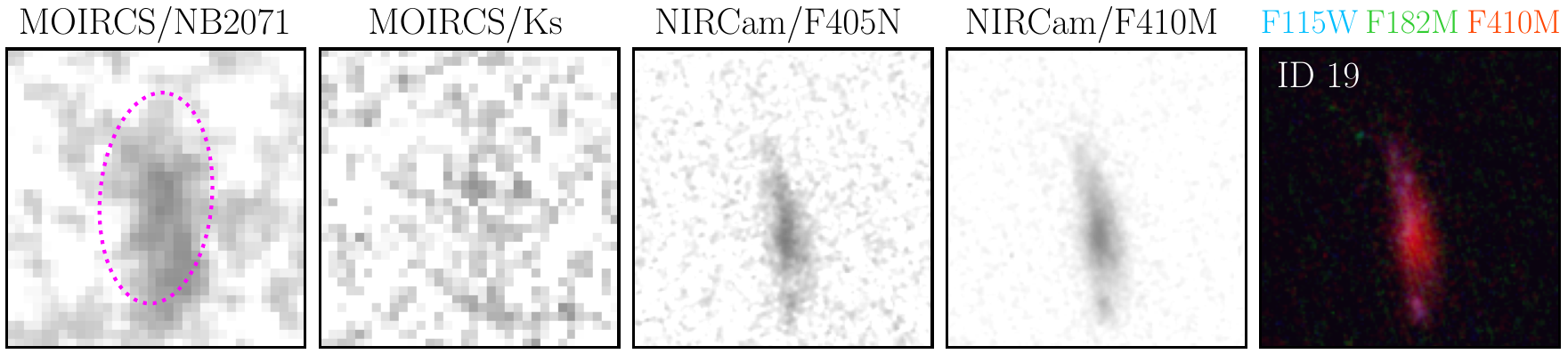}~
\includegraphics[width=0.5\linewidth]{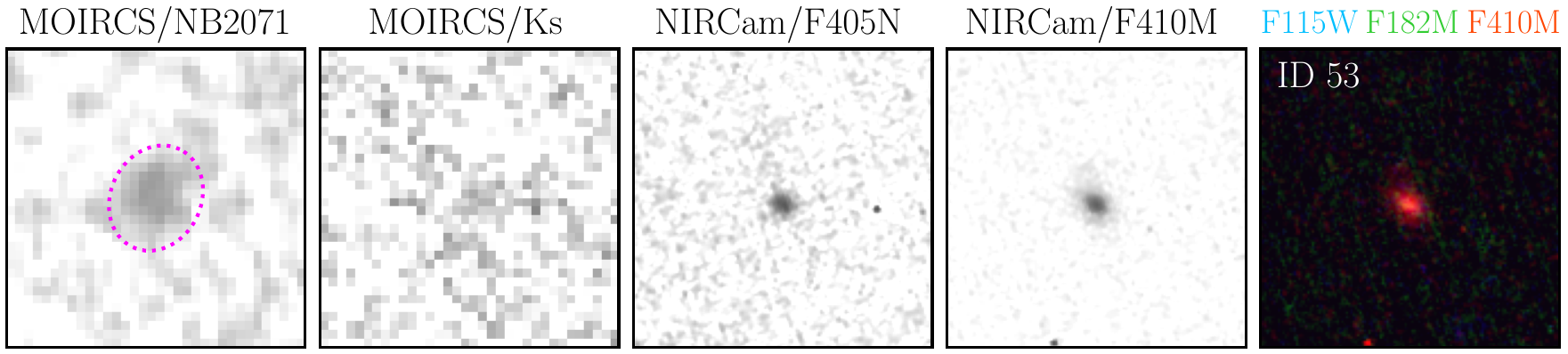}\par
\includegraphics[width=0.5\linewidth]{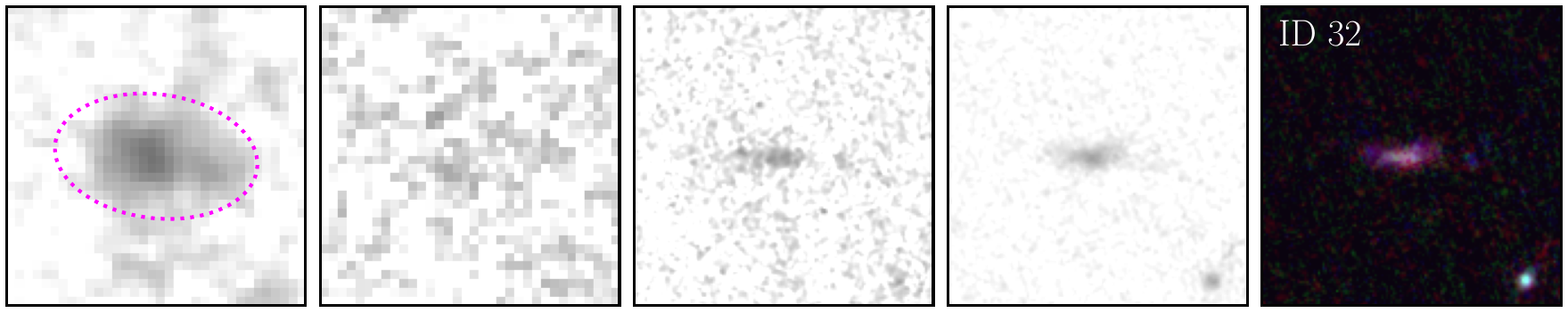}~
\includegraphics[width=0.5\linewidth]{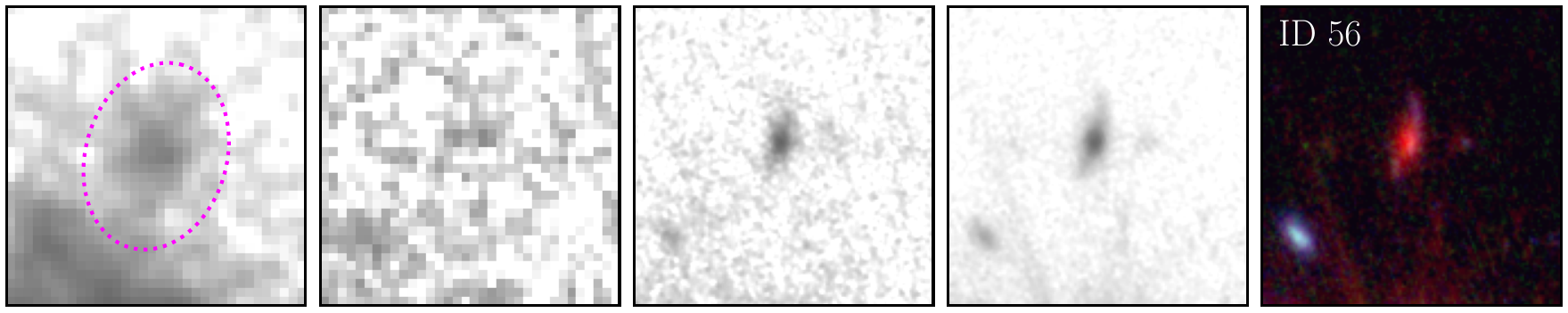}\par
\includegraphics[width=0.5\linewidth]{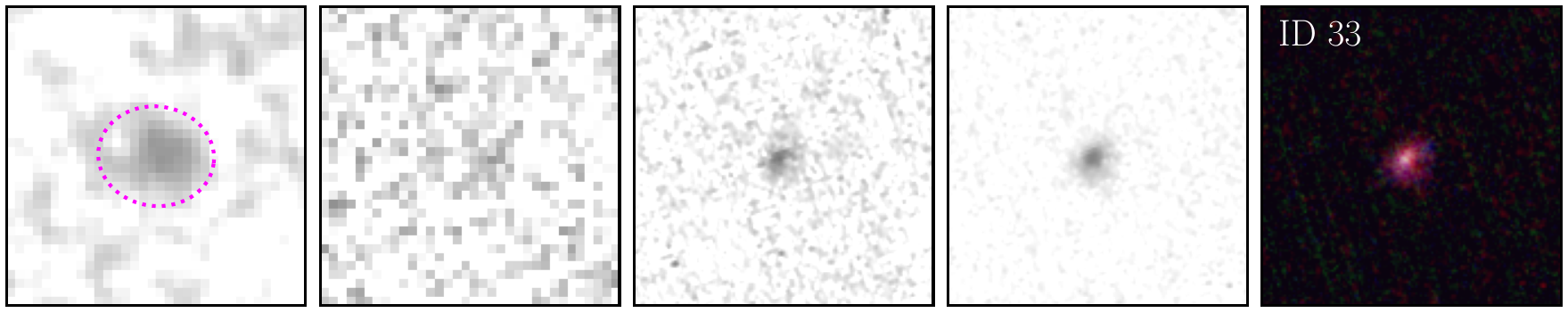}~
\includegraphics[width=0.5\linewidth]{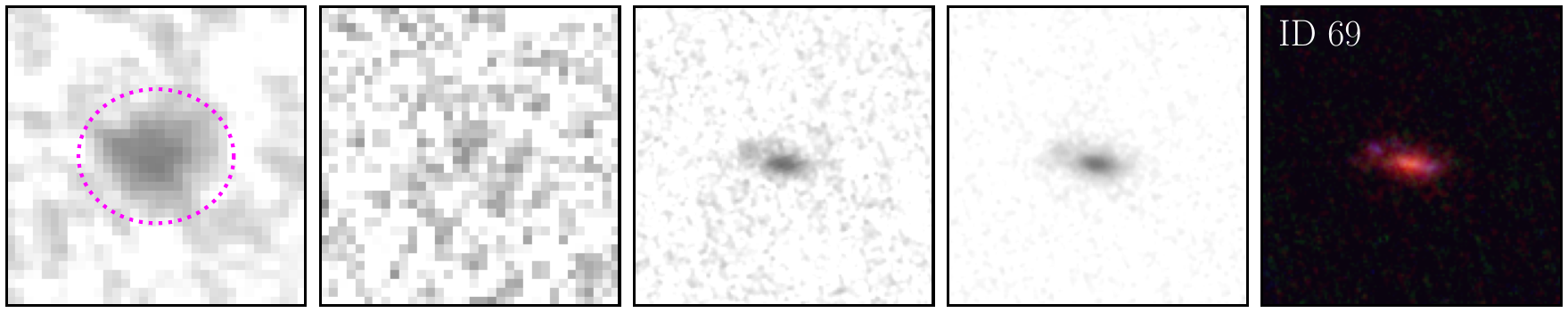}\par
\includegraphics[width=0.5\linewidth]{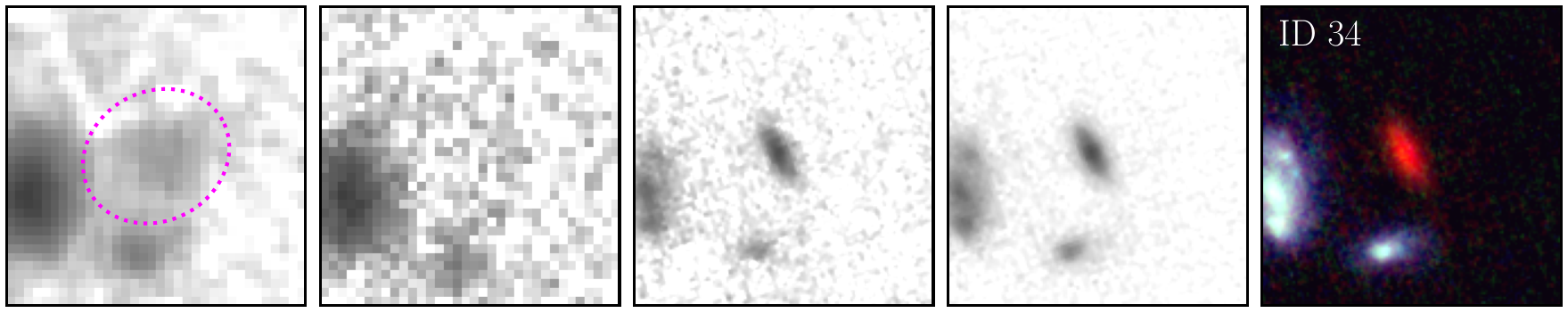}~
\includegraphics[width=0.5\linewidth]{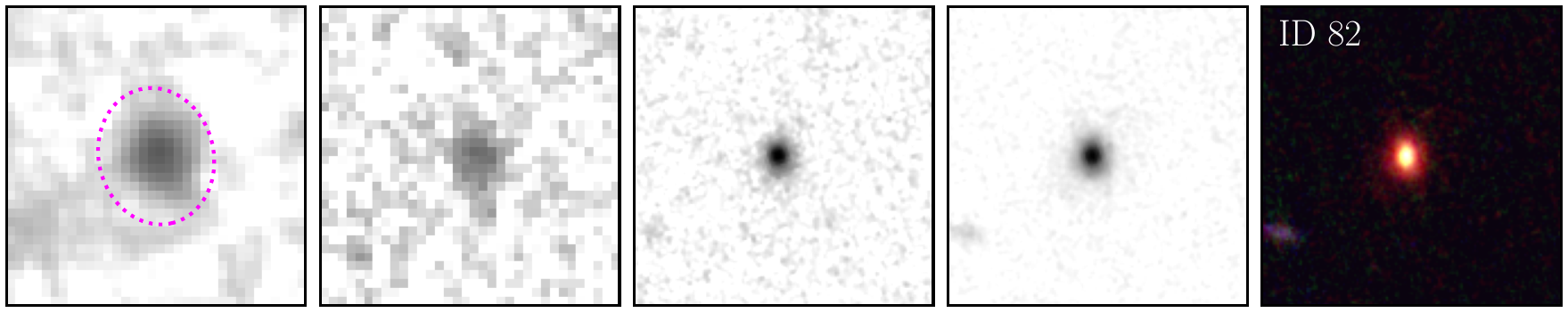}\par
\includegraphics[width=0.5\linewidth]{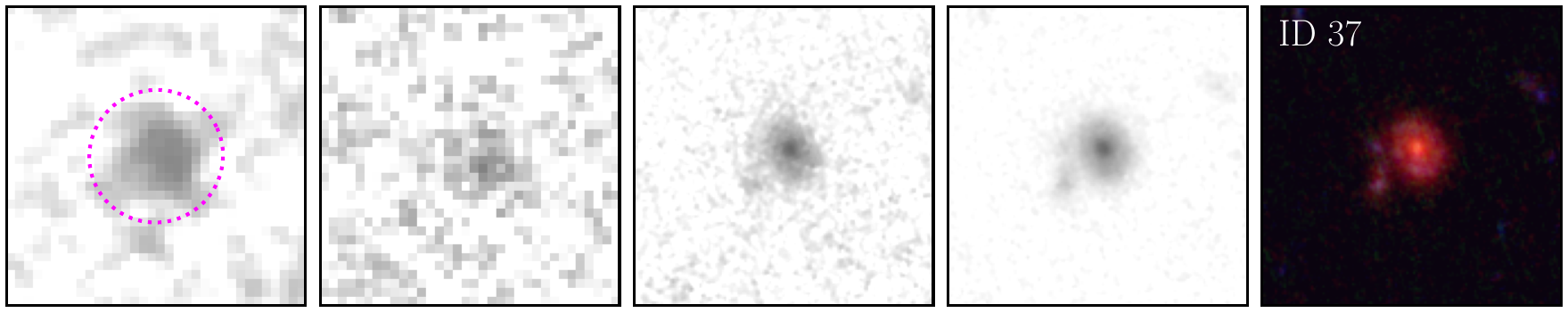}~
\includegraphics[width=0.5\linewidth]{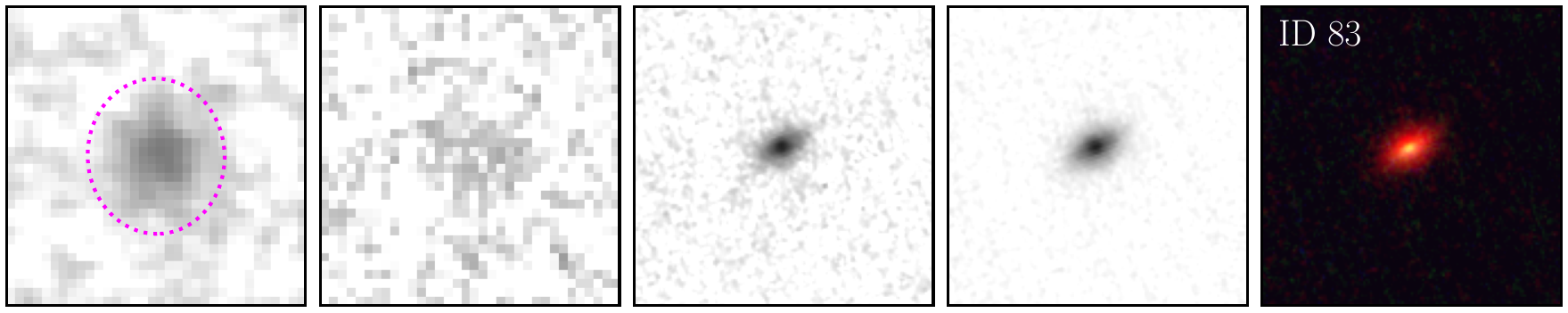}\par
\includegraphics[width=0.5\linewidth]{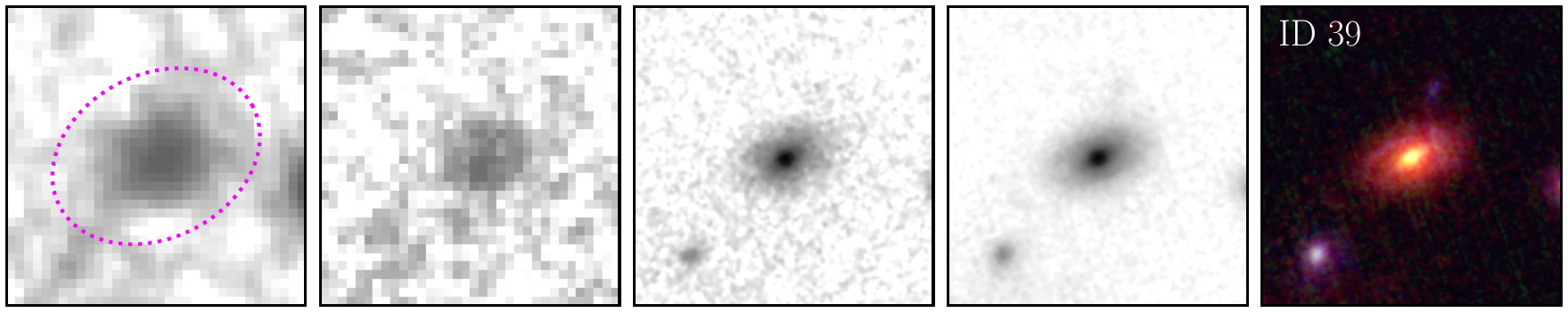}~
\includegraphics[width=0.5\linewidth]{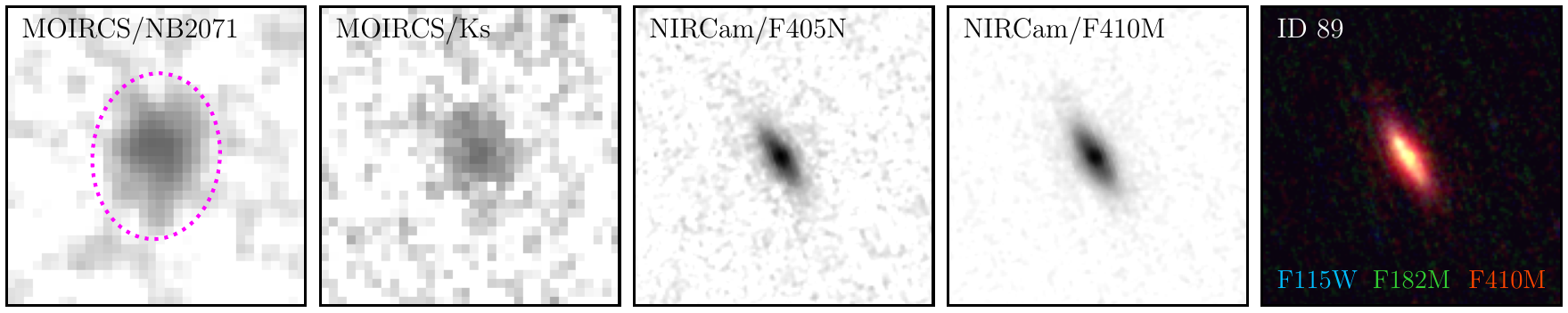}\par
\includegraphics[width=0.5\linewidth]{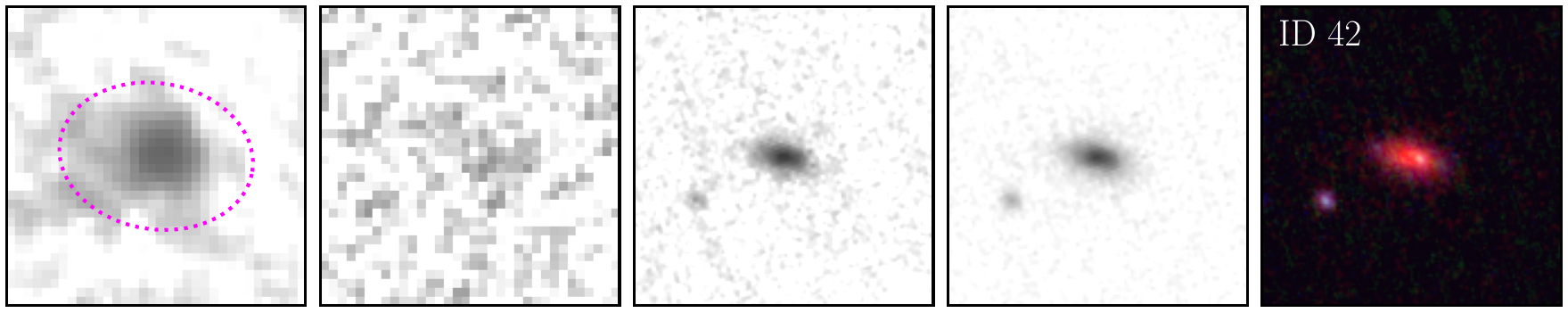}
\caption{Gallery depicting the images of the HAEs without spectroscopic redshift quoted in Table\,\ref{T:appendix}. Sources are ordered by ID in two blocks made of 5 columns each displaying squared cutouts ($\mathrm{4\arcsec\times4\arcsec}$) for each HAE.  Images and symbols follow the scheme outlined in Fig.\,\ref{F:Gallery1}.}
\label{F:Gallery2}
\end{figure*}

\begin{figure*}
\includegraphics[width=0.5\linewidth]{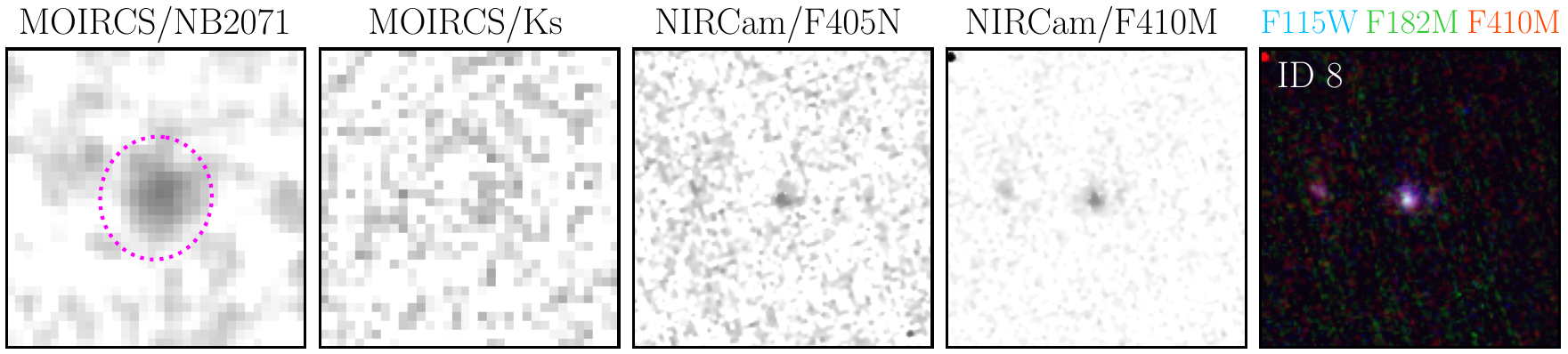}~
\includegraphics[width=0.5\linewidth]{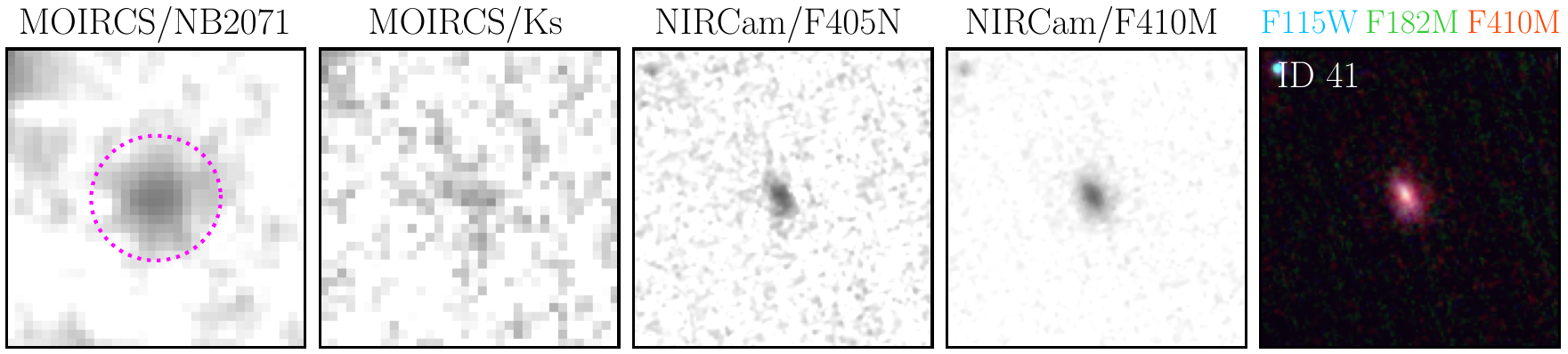}\par
\includegraphics[width=0.5\linewidth]{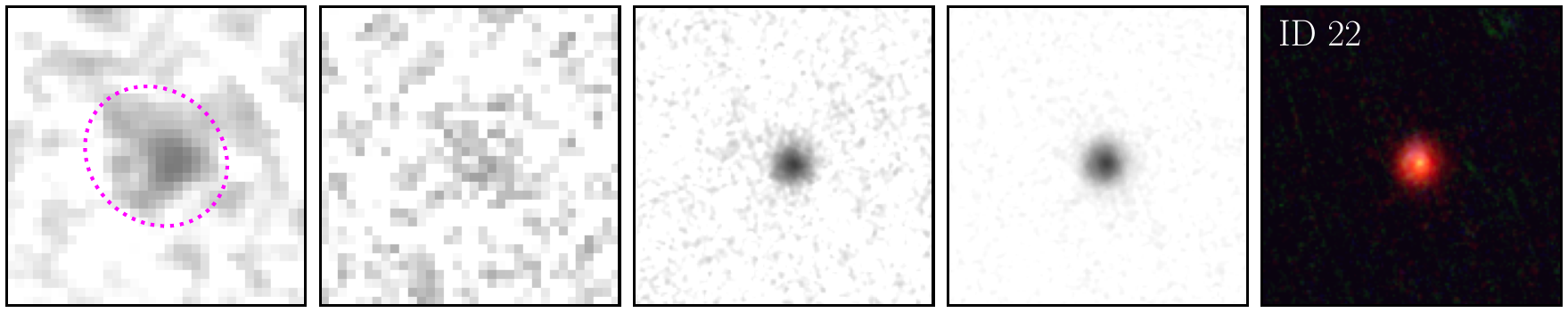}~
\includegraphics[width=0.5\linewidth]{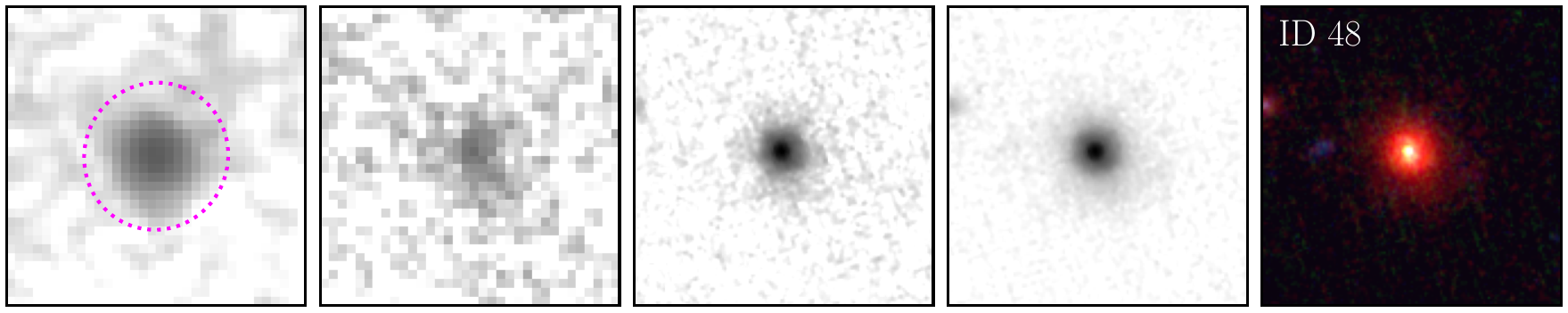}\par
\includegraphics[width=0.5\linewidth]{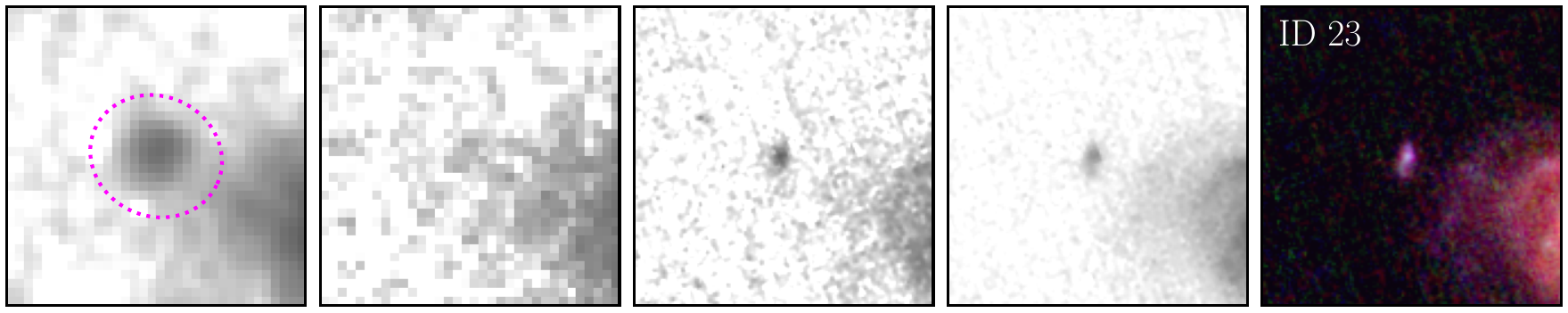}~
\includegraphics[width=0.5\linewidth]{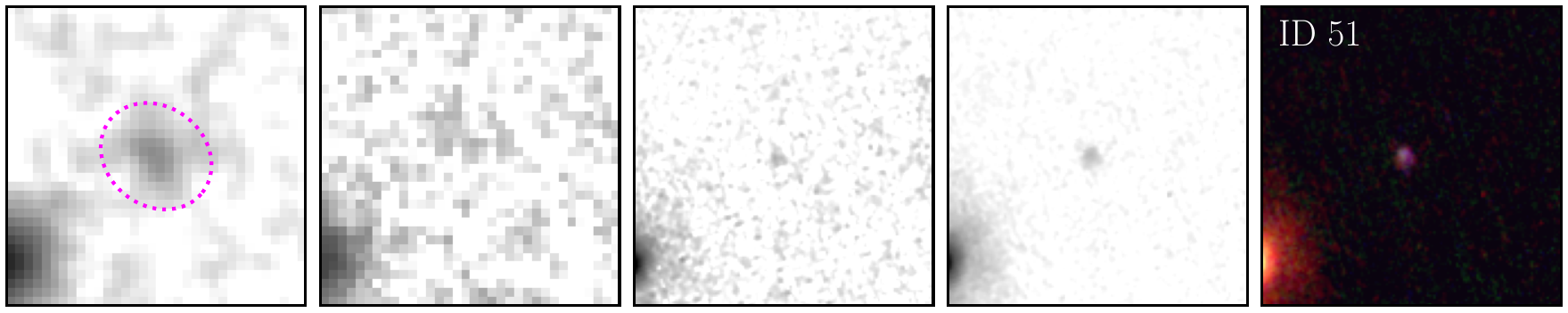}\par
\includegraphics[width=0.5\linewidth]{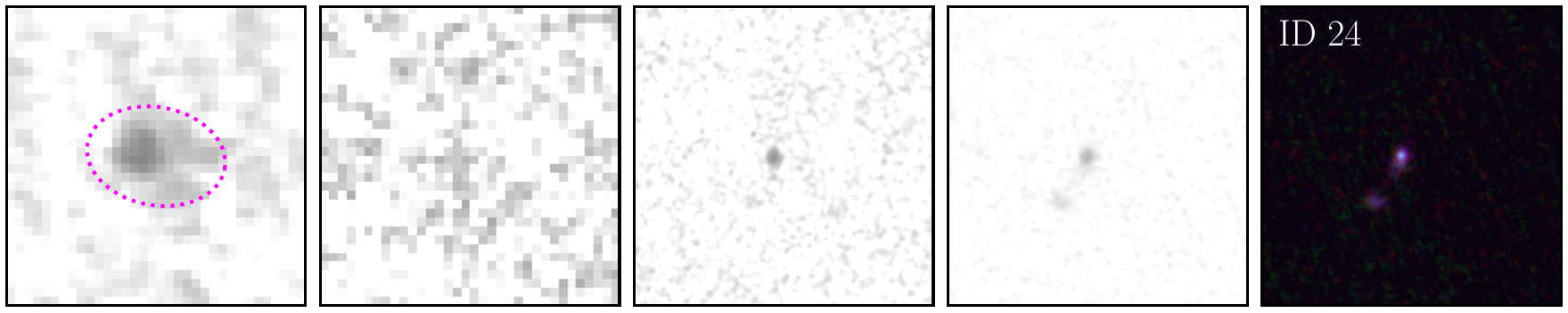}~
\includegraphics[width=0.5\linewidth]{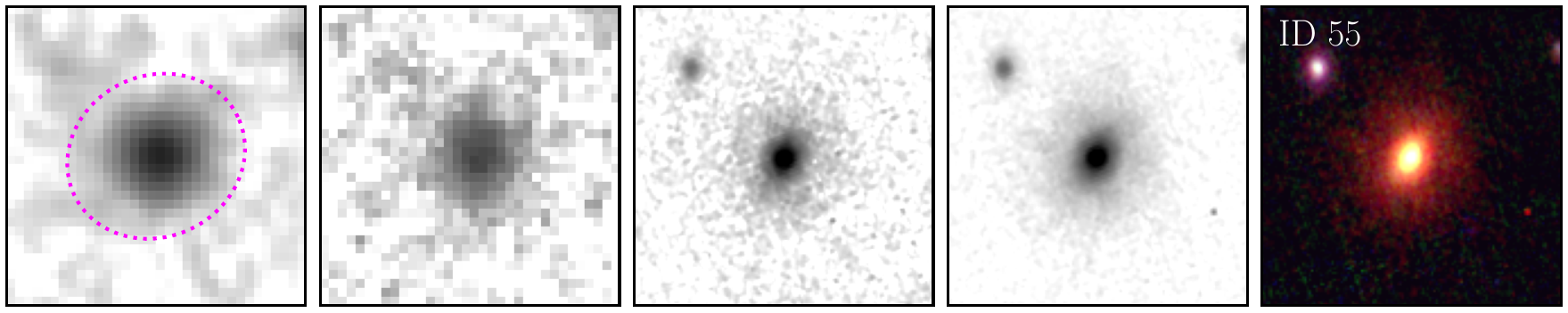}\par
\includegraphics[width=0.5\linewidth]{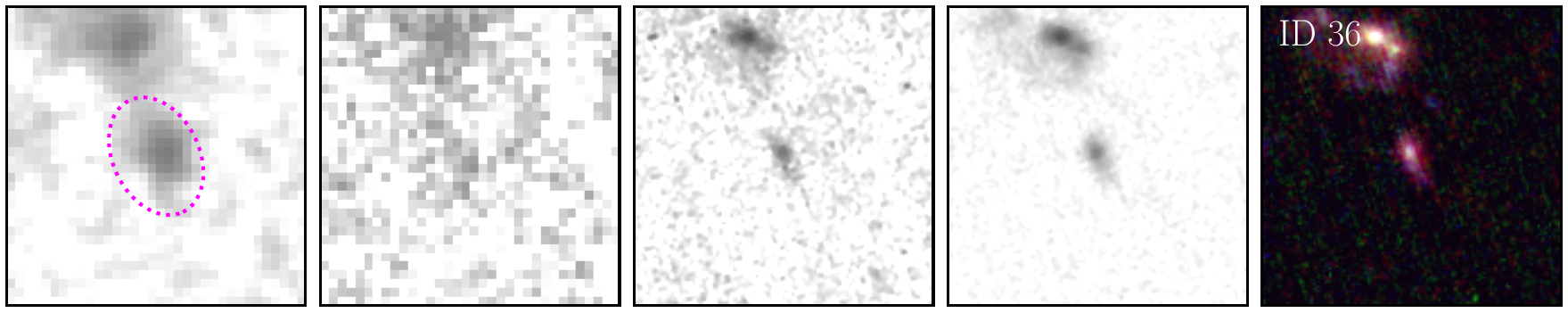}~
\includegraphics[width=0.5\linewidth]{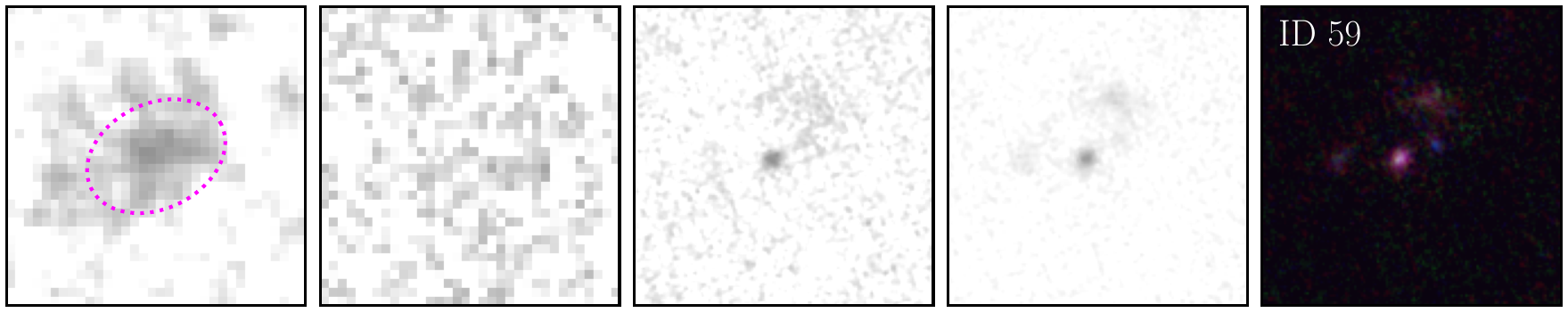}\par
\includegraphics[width=0.5\linewidth]{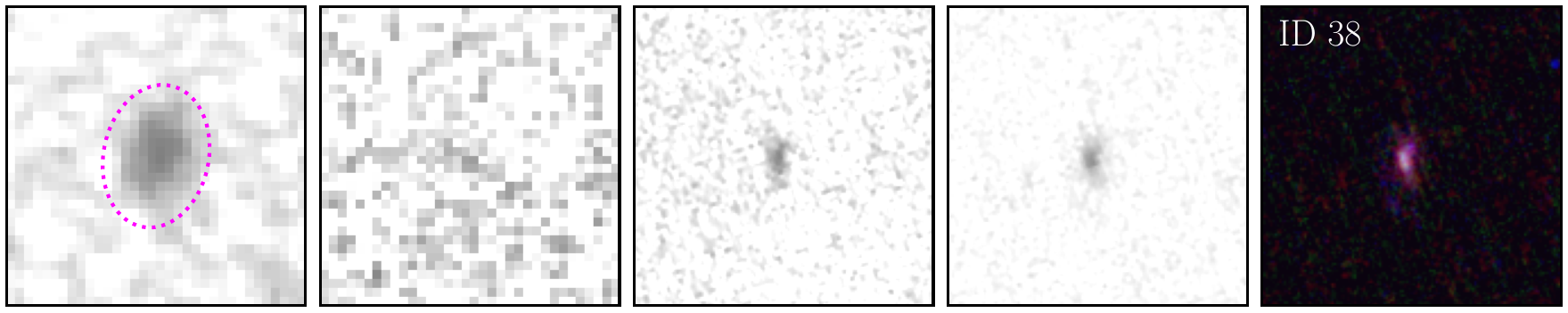}~
\includegraphics[width=0.5\linewidth]{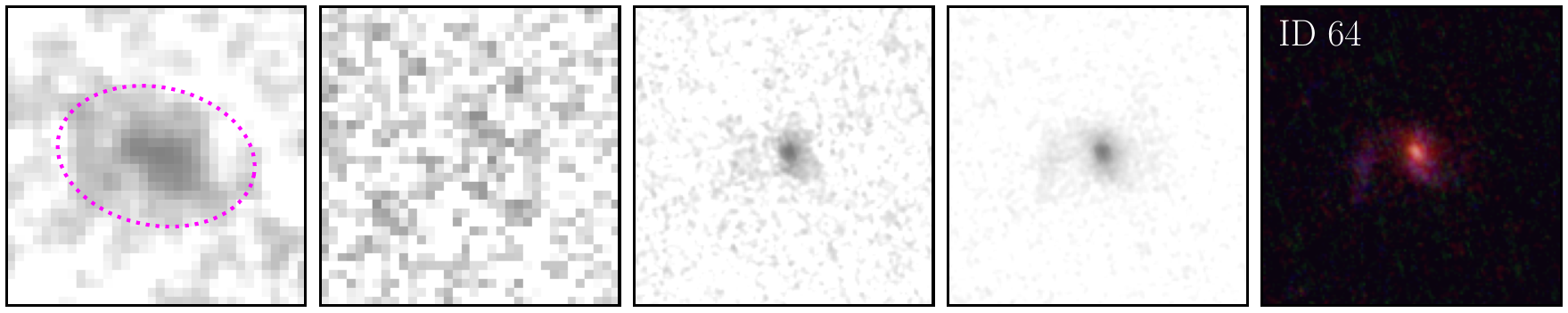}\par
\includegraphics[width=0.5\linewidth]{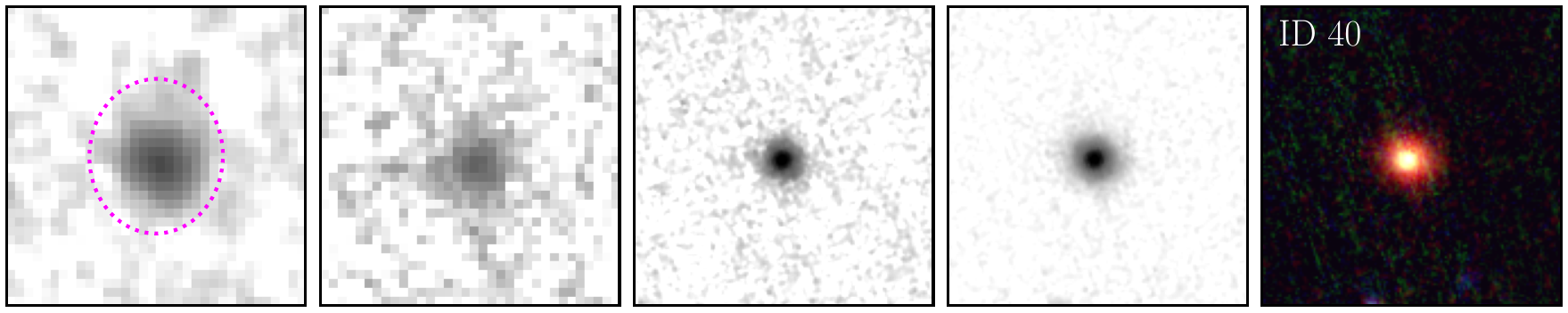}~
\includegraphics[width=0.5\linewidth]{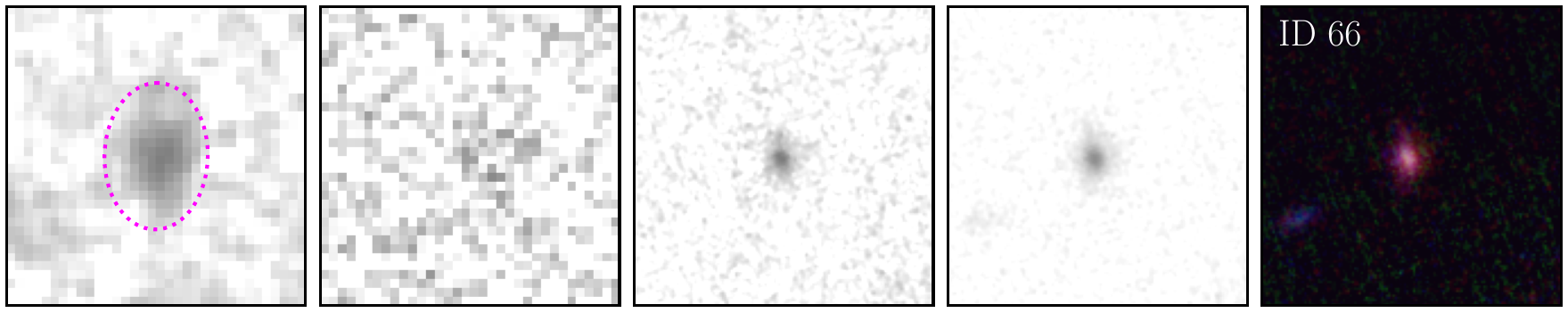}\par
\includegraphics[width=0.5\linewidth]{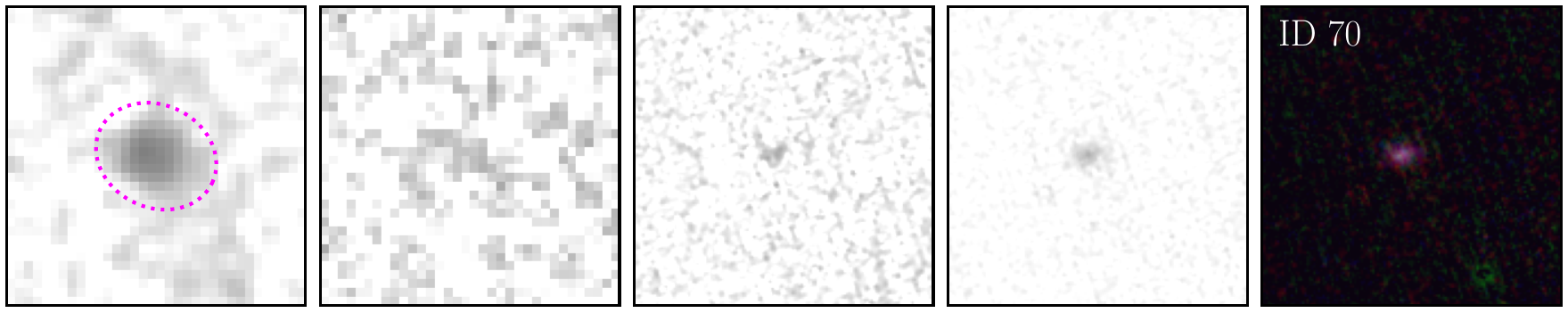}~
\includegraphics[width=0.5\linewidth]{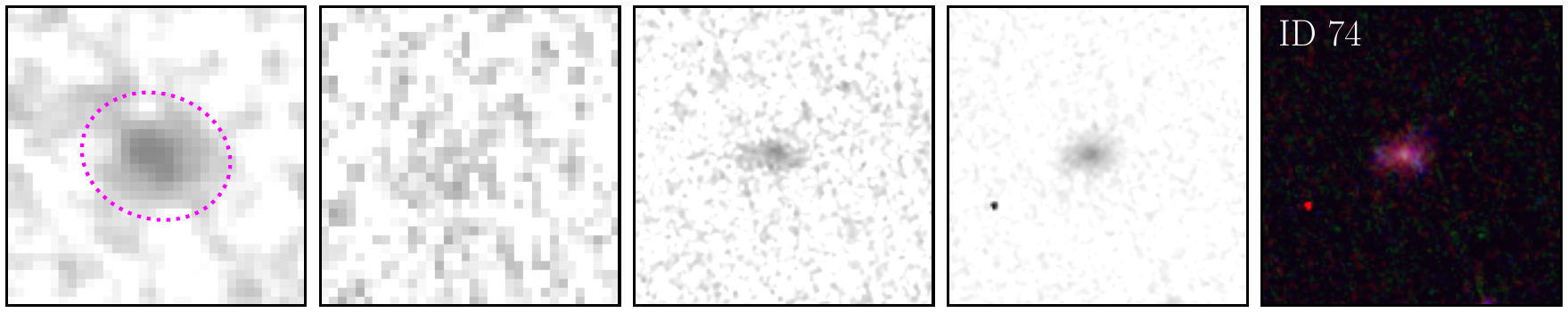}\par
\includegraphics[width=0.5\linewidth]{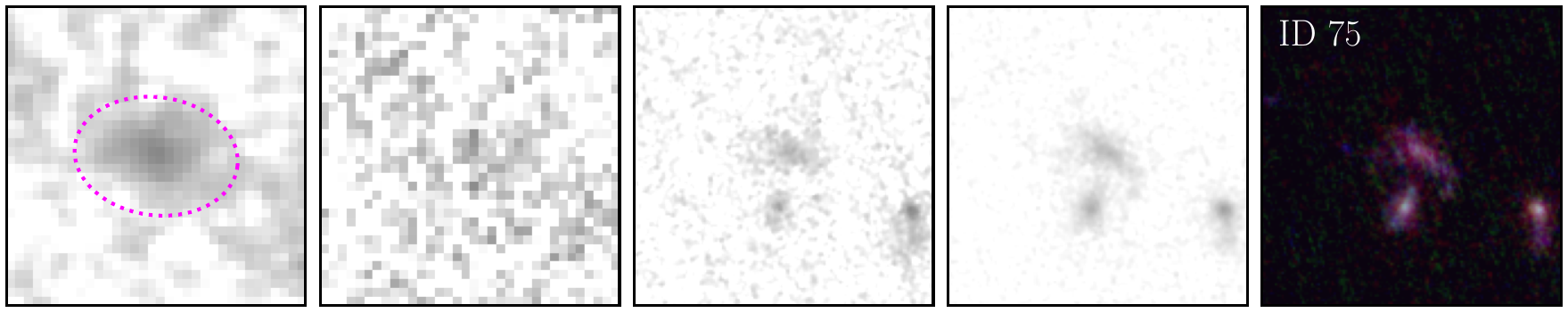}~
\includegraphics[width=0.5\linewidth]{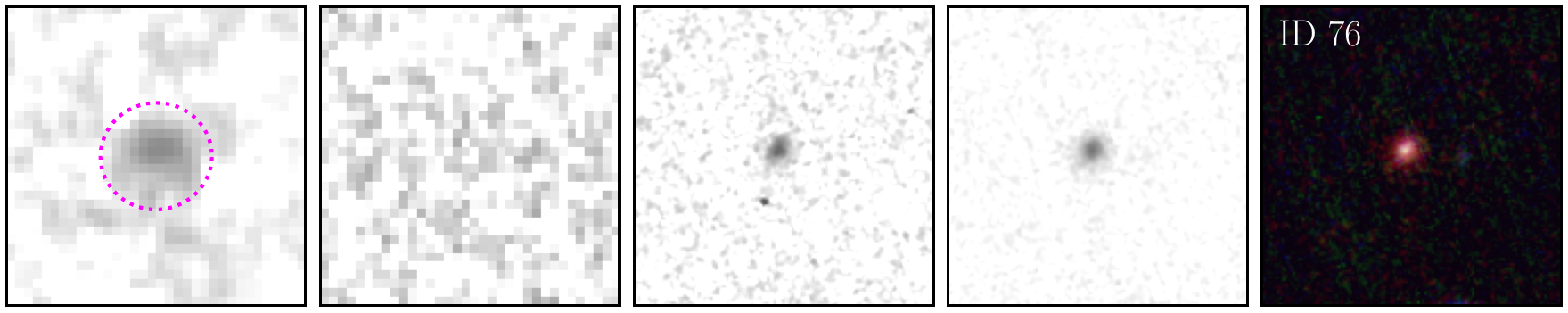}\par
\includegraphics[width=0.5\linewidth]{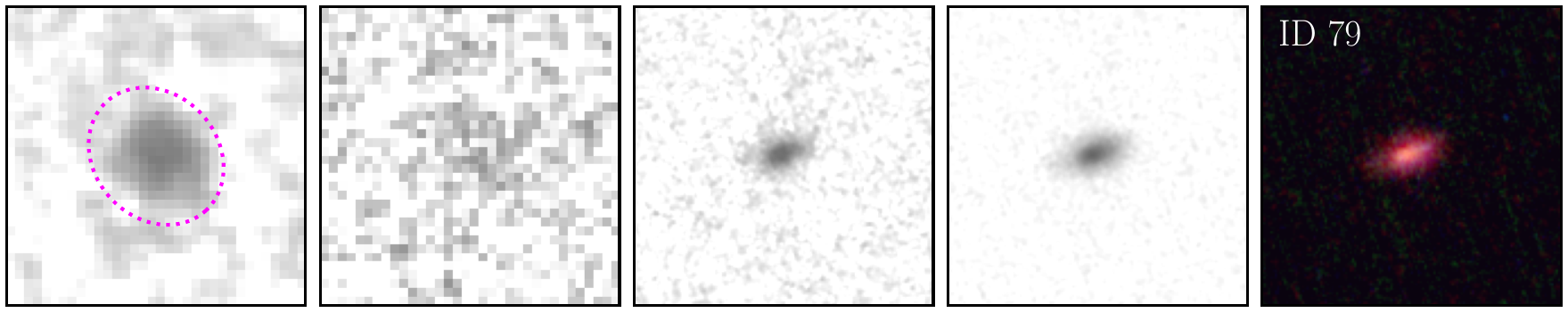}~
\includegraphics[width=0.5\linewidth]{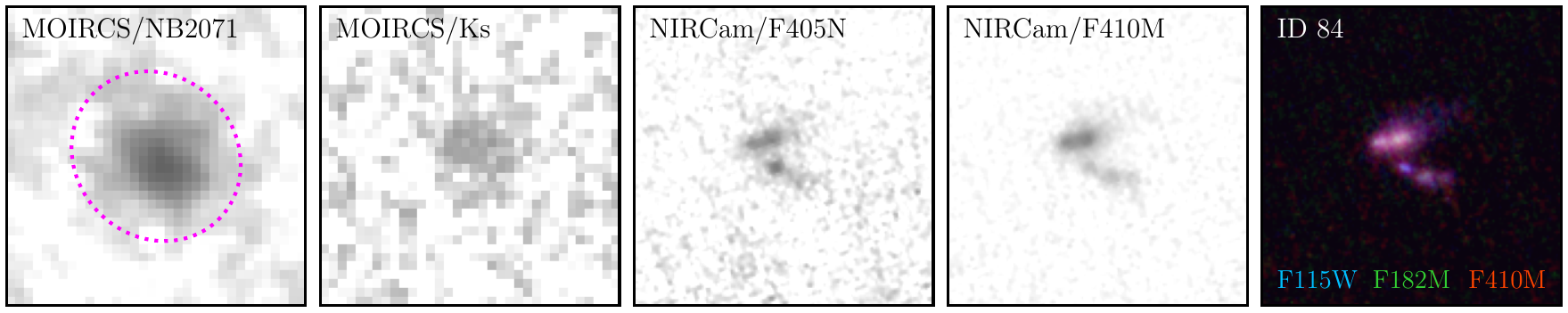}\par
\includegraphics[width=0.5\linewidth]{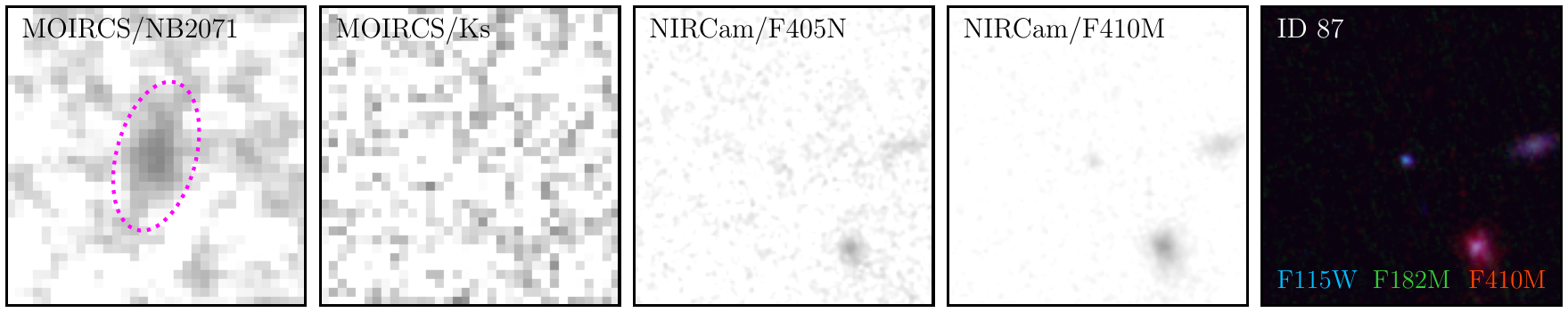}~
\includegraphics[width=0.5\linewidth]{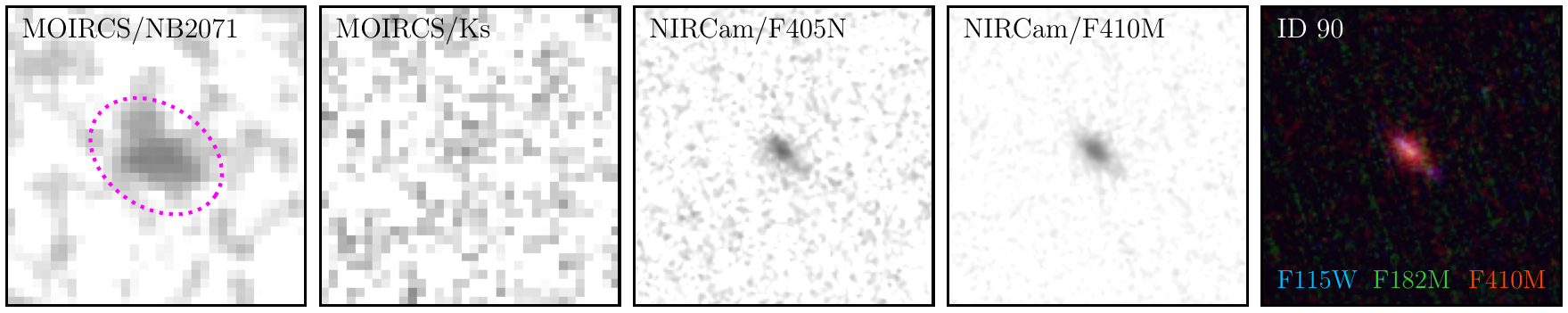}\par
\includegraphics[width=0.5\linewidth]{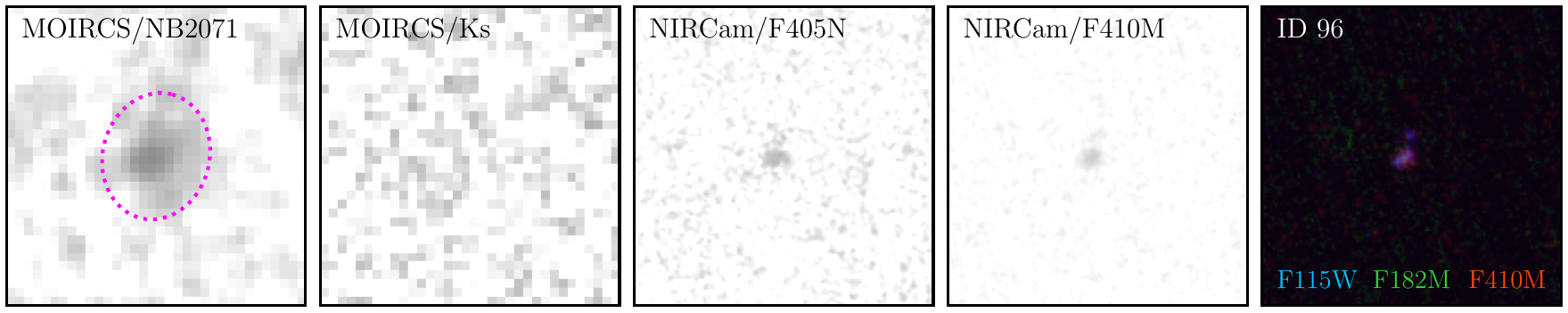}
\caption{Gallery depicting the images of the HAEs discarded due to their lack of $\mathrm{Pa\beta}$ emission in Sect.\ref{SS:EL} or due to their low S/N (Sect.\,\ref{SS:DustProp}). Sources are ordered by ID in two blocks made of 5 columns each displaying squared cutouts ($\mathrm{4\arcsec\times4\arcsec}$) for each HAE. Images and symbols follow the same scheme outlined in Fig.\,\ref{F:Gallery1}.}
\label{F:Gallery3}
\end{figure*}

\section{Summary table}
\begin{table*}
\caption{Main properties of the final sample of HAEs in this work. HAEs with spectroscopic redshift are shown at the start of the table while HAEs without such information are quoted after the horizontal line. IDs, stellar masses, and spectroscopic redshifts are reproduced from \cite{Shimakawa24}. We omit the redshift value for HAEs without spectroscopic confirmation. Extinction values ($\mathrm{A_{H\alpha}}$) are based on the $\mathrm{H\alpha/Pa\beta}$ ratios while star formation rates relate on the latter emission line and dust correction. The PBE flag highlights galaxies fulfilling the selection criteria of this population in S24. The X-ray emission flag corresponds to objects selected by \cite{Tozzi22a}. The SMG flag identifies the crossmatch of our sample with the very dusty objects reported in \cite{Dannerbauer14}.}
\centering
\begin{tabular}{cccccccc}
\hline
\noalign{\vskip 0.1cm}
ID &  z & $\mathrm{A_{H\alpha}}$ & $\mathrm{\log SFR_{Pa\beta}/M_\odot\,yr^{-1}}$ & $\mathrm{\log M_*/M_\odot}$ & PBE & X-ray & SMG \\ 
\noalign{\vskip 0.1cm}
   &    &      (mag)  & ($\mathrm{M_\odot\,yr^{-1}}$) & ($\mathrm{M_\odot}$) &  & \\ 
\noalign{\vskip 0.1cm}
\hline 
9 & 2.1437 & $2.54\pm0.68$ & $2.13\pm0.12$ & $10.61^{+0.09}_{-0.11}$ & 0  & 0 & - \\[0.1cm]  
11 & 2.1627 & $0.00\pm1.17$ & $0.86\pm0.26$ & $10.00^{+0.07}_{-0.08}$ & 1  & 0 & - \\[0.1cm]
16 & 2.1577 & $1.11\pm0.58$ & $1.75\pm0.10$ & $9.91^{+0.05}_{-0.06}$ & 1  & 0 & - \\[0.1cm] 
17 & 2.1617 & $2.50\pm0.88$ & $1.90\pm0.11$ & $10.11^{+0.06}_{-0.07}$ & 0  & 0 & - \\[0.1cm]
18 & 2.1609 & $2.04\pm0.73$ & $1.80\pm0.08$ & $9.86^{+0.12}_{-0.17}$ & 0 & 0 & - \\[0.1cm] 
21 & 2.1575 & $2.18\pm0.66$ & $1.93\pm0.09$ & $10.03^{+0.10}_{-0.12}$ & 1  & 0 & - \\[0.1cm] 
28 & 2.1532 & $1.94\pm1.03$ & $1.64\pm0.19$ & $10.88^{+0.16}_{-0.25}$ & 0  & 1 & - \\[0.1cm]
30 & 2.1513 & $2.30\pm0.42$ & $2.44\pm0.06$ & $10.94^{+0.02}_{-0.02}$ & 0  & 0 & - \\[0.1cm]
35 & 2.1551 & $0.95\pm0.48$ & $1.80\pm0.08$ & $10.31^{+0.04}_{-0.04}$ & 1  & 0 & - \\[0.1cm] 
45 & 2.1415 & $2.35\pm0.51$ & $2.35\pm0.09$ & $9.59^{+0.07}_{-0.09}$ & 0  & 0 & - \\[0.1cm] 
46 & 2.1557 & $0.69\pm0.30$ & $2.24\pm0.06$ & $10.32^{+0.41}_{-0.40}$ & 0  & 1 & - \\[0.1cm] 
49 & 2.1661 & $0.75\pm0.72$ & $1.32\pm0.10$ & $9.46^{+0.12}_{-0.16}$ & 0  & 0 & - \\[0.1cm]
54 & 2.1480 & $0.38\pm0.69$ & $1.68\pm0.16$ & $11.21^{+0.02}_{-0.02}$ & 0  & 0 & - \\[0.1cm] 
57 & 2.1523 & $0.36\pm1.18$ & $0.91\pm0.24$ & $9.40^{+0.06}_{-0.08}$ & 0  & 0 & - \\[0.1cm] 
58 & 2.1568 & $1.89\pm0.67$ & $1.92\pm0.11$ & $10.83^{+0.07}_{-0.08}$ & 0  & 1 & - \\[0.1cm]
60 & 2.1634 & 0 & $0.87\pm0.30$ & $10.15^{+0.09}_{-0.11}$ & 0  & 0 & - \\[0.1cm] 
65 & 2.1628 & $1.14\pm0.57$ & $1.67\pm0.08$ & $9.96^{+0.13}_{-0.18}$ & 1  & 0 & - \\[0.1cm]
67 & 2.1634 & 0 & $0.90\pm0.23$ & $9.68^{+0.06}_{-0.07}$ & 0  & 0 & - \\[0.1cm]
71 & 2.1630 & $1.08\pm0.58$ & $1.71\pm0.11$ & $10.51^{+0.13}_{-0.18}$ & 1  & 1 & - \\[0.1cm]
73\tablenotemark{a} & 2.1618 & $2.97\pm0.07$ & $3.88\pm0.01$ & $12.42^{+0.06}_{-0.07}$ & 1  & 1 & DKB07 \\[0.1cm]
80 & 2.1606 & $0.53\pm0.78$ & $1.56\pm0.17$ & $10.68^{+0.07}_{-0.08}$ & 0  & 0 & - \\[0.1cm]
93 & 2.1516 & $1.56\pm0.76$ & $1.78\pm0.15$ & $10.39^{+0.16}_{-0.26}$ & 0  & 0 & DKB15 \\[0.1cm]
\hline
\noalign{\vskip 0.1cm}
19 & - & $0.15\pm0.78$ & $1.21\pm0.15$ & $9.74^{+0.08}_{-0.10}$ & 0  & 0 & - \\[0.1cm]
32 & - & $0$ & $1.01\pm0.21$ & $9.69^{+0.06}_{-0.07}$ & 0  & 0 & - \\[0.1cm]
33 & - & $2.76\pm0.80$ & $1.90\pm0.08$ & $9.66^{+0.10}_{-0.13}$ & 0  & 0 & - \\[0.1cm]
34 & - & $0$ & $1.24\pm0.23$ & $9.97^{+0.15}_{-0.22}$ & 0  & 0 & - \\[0.1cm]
37 & - & $1.94\pm0.88$ & $1.58\pm0.10$ & $10.26^{+0.12}_{-0.17}$ & 0  & 0 & - \\[0.1cm]
39 & - & $1.01\pm0.66$ & $1.71\pm0.13$ & $10.71^{+0.10}_{-0.13}$ & 0  & 0 & - \\[0.1cm]
42 & - & $0.44\pm0.57$ & $1.49\pm0.1$ & $9.90^{+0.12}_{-0.17}$ & 1  & 0 & - \\[0.1cm]
53 & - & $1.78\pm0.97$ & $1.56\pm0.17$ & $9.83^{+0.19}_{-0.33}$ & 1  & 0 & - \\[0.1cm]
56 & - & $0$ & $0.98\pm0.26$ & $9.76^{+0.18}_{-0.32}$ & 0  & 0 & - \\[0.1cm]
69 & - & $1.36\pm0.76$ & $1.51\pm0.10$ & $10.02^{+0.11}_{-0.14}$ & 1  & 0 & - \\[0.1cm]
82 & - & $0.95\pm1.16$ & $1.17\pm0.22$ & $10.59^{+0.07}_{-0.08}$ & 0  & 0 & - \\[0.1cm]
83 & - & $0$ & $1.00\pm0.22$ & $10.38^{+0.16}_{-0.27}$ & 0  & 1 & - \\[0.1cm]
89 & - & $1.45\pm1.13$ & $1.44\pm0.23$ & $10.69^{+0.03}_{-0.04}$ & 0  & 0 & - \\[0.1cm]
\hline
\end{tabular}
\tablenotetext{a}{Spiderweb radio galaxy.}
\label{T:appendix}
\end{table*}

%

\facilities{JWST (NIRCam), Subaru (MOIRCS)}


\software{Astropy \citep{Astropy13, Astropy18}, 
          Photutils \citep{Photutils23},  
          Source Extractor \citep{Bertin96},
          PSFEx \citep{Bertin11}.
          }
          

\bibliography{sample631}{}
\bibliographystyle{aasjournal}



\end{document}